\documentclass[preprint]{aastex}

\usepackage{times}

\usepackage{amssymb}
\usepackage{amsbsy}

\usepackage{amsmath}
\usepackage{latexsym}   

\usepackage{psfig}
\psfigurepath{./figures/}
\pssilent
\usepackage{natbib}
\newcommand{\glh}{\mbox{$ \hat \lambda$}}

\newcommand{\gs}{\sigma}

\def\bm#1{\mbox{\boldmath $#1$}}

\newcommand{\mgG}{\mbox{$\bm \Gamma$}}

\newcommand{\mLambda}{\mbox{$\bm \Lambda$}}

\newcommand{\bsigma}{{\bm\sigma}}

\newcommand{\rH}{^{ \raisebox{1pt}{$\rm \scriptscriptstyle H$}}}
\newcommand{\rT}{^{ \raisebox{1.2pt}{$\rm \scriptstyle T$}}}

\newcommand{\rE}{{\rm E}}

\newcommand{\beq}{\begin{equation}}
\newcommand{\enq}{\end{equation}}

\newcommand{\bea}{\begin{array}}
\newcommand{\ena}{\end{array}}
\newcommand{\bds}{\begin {itemize}}
\newcommand{\eds}{\end {itemize}}

\newcommand{\larrow}{{\larrow}}

\newcommand{\ie}{\hbox{i.e., }}
\newcommand{\eg}{\hbox{e.g., }}

\newcommand{\viz}{\hbox{viz.\ }}

\newcommand{\INR}{\ensuremath{\mathrm{INR}}}

\newcommand{\MDL}{\ensuremath{\mathrm{MDL}}}
\newcommand{\vect}{\ensuremath{\mathrm{vec}}}

\newcommand{\spann}{\ensuremath{\mathrm{span}}}

\newcommand{\cC}{\ensuremath{\mathcal{C}}}

\newcommand{\cH}{\ensuremath{\mathcal{H}}}

\newcommand{\cN}{\ensuremath{\mathcal{N}}}

\newcommand{\va}{\ensuremath{\mathbf{a}}}

\newcommand{\vecr}{\ensuremath{\mathbf{r}}}
\newcommand{\vs}{\ensuremath{\mathbf{s}}}

\newcommand{\vx}{\ensuremath{\mathbf{x}}}

\newcommand{\ba}{\ensuremath{\mathbf{a}}}
\newcommand{\bah}{\ensuremath{\mathbf{\hat{a}}}}

\newcommand{\br}{\ensuremath{\mathbf{r}}}
\newcommand{\bs}{\ensuremath{\mathbf{s}}}
\newcommand{\bu}{\ensuremath{\mathbf{u}}}

\newcommand{\bx}{\ensuremath{\mathbf{x}}}

\newcommand{\mA}{\ensuremath{\mathbf{A}}}

\newcommand{\mB}{\ensuremath{\mathbf{B}}}

\newcommand{\mC}{\ensuremath{\mathbf{C}}}

\newcommand{\mI}{\ensuremath{\mathbf{I}}}

\newcommand{\mL}{\ensuremath{\mathbf{L}}}
\newcommand{\mLb}{\ensuremath{\mathbf{\bar{L}}}}

\newcommand{\mR}{\ensuremath{\mathbf{R}}}
\newcommand{\mRh}{\ensuremath{\mathbf{\hat{R}}}}
\newcommand{\mRt}{\ensuremath{\mathbf{\tilde{R}}}}

\newcommand{\mU}{\ensuremath{\mathbf{U}}}

\newcommand{\mX}{\ensuremath{\mathbf{X}}}

\newcommand{\bC}{\ensuremath{\mathbf{C}}}

\newcommand{\bP}{\ensuremath{\mathbf{P}}}

\newcommand{\bR}{\ensuremath{\mathbf{R}}}

\newcommand{\bX}{\ensuremath{\mathbf{X}}}

\newcommand{\DL}{\begin{dashlist}}
\newcommand{\DLE}{\end{dashlist}}

\shorttitle{Interference mitigation in radio astronomy}
\shortauthors{Leshem, Van der Veen and Boonstra}

\begin{document}

\title{MULTICHANNEL INTERFERENCE MITIGATION TECHNIQUES IN RADIO ASTRONOMY}
\author{Amir Leshem,  Alle-Jan van der Veen}
\affil{
    Information Technology and Systems\\
    Delft University of Technology\\ 
    2628 CD Delft, The Netherlands\\
    email: leshem@cas.et.tudelft.nl, allejan@cas.et.tudelft.nl
}

\email{leshem@cas.et.tudelft.nl,allejan@cas.et.tudelft.nl}
\and
\author{Albert-Jan Boonstra}
\affil{
    NFRA/ASTRON\\
    Postbus 2, 7990 AA Dwingeloo, The Netherlands\\
    e-mail:boonstra@nfra.nl
}
\email{boonstra@nfra.nl}

\begin{abstract}
    Radio-astronomical observations are increasingly corrupted by RF
    interference, and online detection and filtering algorithms are
    becoming essential. To facilitate the introduction of such
    techniques into radio astronomy, we formulate the astronomical
    problem in an array signal processing language, and give an
    introduction to some elementary algorithms from that field. We
    consider two topics in detail: interference detection by rank
    estimation of short-term covariance matrices, and spatial filtering
    by subspace estimation and projection.  We discuss experimental
    data collected at the Westerbork radio telescope, and illustrate
    the effectiveness of the space-time detection and blanking process
    on the recovery of a 3C48 absorption line in the presence of GSM
    mobile telephony interference.
\end{abstract}

\keywords{methods: statistical; instrumentation: interferometers; methods: analytical} 

\clearpage
\section{INTRODUCTION}

    Radio-astronomical observations are increasingly corrupted by RF
    interferers such as wireless communication and satellite navigation
    signals. Online detection and filtering algorithms are essential to
    reduce the effect of interference to an acceptable level. However,
    existing methods have a limited scope. Until now, the most widely
    implemented algorithm is a single-channel total power change
    detector, followed by a blanking of the correlator output.
    \citet{friedman96a} has implemented an improved power detector at
    the RATAN600, based on detection of change in the power.
    \citet{weber97} proposed the use of the quantized correlation at
    all lags to test the presence of interference.  Another detector
    based on wavelet decomposition has been proposed by
    \citet{maslakovic96}.  These are all single channel detectors which
    do not exploit the spatial properties of the interference.  The
    only detector which considered combining multiple telescopes for
    improved detection and blanking was proposed by \citet{kasper82}
    for low frequency interferometry, where a robust data censoring
    method based on the temporal behavior of the cross spectrum was
    proposed. This requires a large number of estimated spectra
    ($10^5$) to obtain reliable robust estimates, and only two channels
    are used. Finally, adaptive filtering techniques have recently been
    considered by \citet{bradley98} who propose to excise interference
    from the Green-Bank radio telescope using a reference antenna and
    an LMS type algorithm.

    Our aim in this paper is to introduce modern array signal
    processing techniques to the context of radio astronomy, and to
    investigate the merits of {\em multichannel} detection and
    filtering algorithms at the Westerbork Synthesis Radio Telescope
    (WSRT). By combining cross-correlation information of a large number
    of sensor pairs, we can increase the detection performance
    significantly, and also estimate the spatial signature of
    interferers. In essence, our approach is to compute (on-line)
    short-term spatial correlation matrices in narrow sub-bands,
    and then to compute the eigenvalue decomposition of each of these
    matrices \citep{leshem99spawc}.
    A rank estimate based on the eigenvalues allows to detect the
    number of interfering signals in each time-frequency slot, and the
    dominant eigenvectors give information on the ``spatial signature''
    of the interferers.

    After detection, we can follow two directions.
    We can reduce the interference by rejecting
    corrupted time-frequency slots (blanking).  This approach is
    suitable for time-slotted communication signals such as the
    European mobile telephony standard GSM, or the TDMA (time-division
    multiple access)-based mobile telephony standards IS-54/136 in the US.

    A more challenging approach is to also use the eigenvector information.
    Indeed, we can project out those dimensions in the spatial correlation 
    matrices that correspond to the spatial signature vectors of the
    interference.  Such spatial filtering techniques will greatly
    enhance the performance of observations with continuously-present
    interference.

    The effectiveness of the space-time detection and blanking process
    is demonstrated by applying the algorithms to data measured at the
    WSRT using an on-line 8-channel recording system.  As will be shown
    in section \ref{sec:experiments}, we were able to recover an
    absorption line of 3C48 which was completely masked by a
    superimposed GSM interference, and could not be recovered by single
    channel techniques.

    The paper is written in a tutorial style, to appeal to both the radio
    astronomy and the signal processing communities.
    The structure of the paper is as follows.  After posing the
    astronomical measurement equations in section \ref{sec:astron}, we
    reformulate the model in terms of array processing matrix language
    in section \ref{sec:array}. We then introduce RF interference and
    describe its effect on the received data.  In section
    \ref{sec:detection} we discuss various detection algorithms.  We
    compare the single and multichannel detectors, for the case of a
    narrow-band interferer with known spatial signature vector, and
    then present two multichannel detectors that do not assume this
    knowledge. We then move to spatial filtering techniques in
    section \ref{sec:spatfilt}, where we formulate the basic ideas and
    describe a projections based approach.  Finally, experimental results
    on multichannel blanking are shown in section \ref{sec:experiments}.

\section{ASTRONOMICAL MEASUREMENT EQUATIONS}		\label{sec:astron}

    In this section we briefly describe a simplified mathematical model for the 
    astronomical measurement process. Our discussion follows the
    introduction in \citet{perley89}.  The purpose of this is to connect
    to a matrix formulation commonly used in array signal processing, in
    the next section.

\begin{figure}
  \begin{center}
	   $(a)$
\begin{picture}(0,0)
\includegraphics{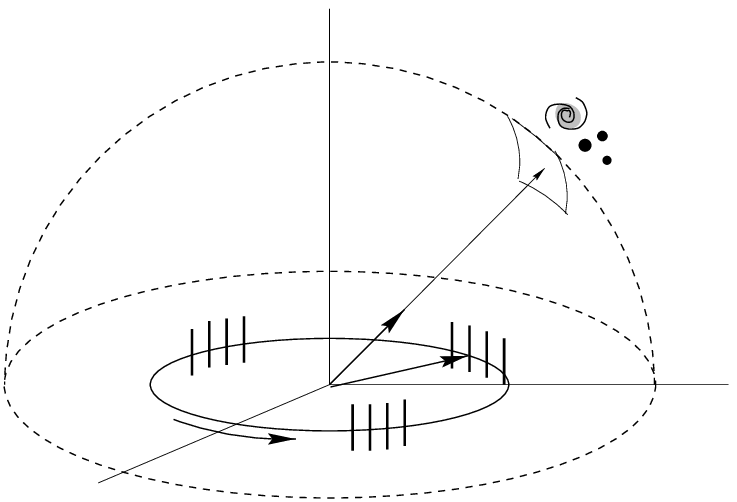}
\end{picture}
\setlength{\unitlength}{1460sp}
\begingroup\makeatletter\ifx\SetFigFont\undefined
\def\x#1#2#3#4#5#6#7\relax{\def\x{#1#2#3#4#5#6}}%
\expandafter\x\fmtname xxxxxx\relax \def\y{splain}%
\ifx\x\y   
\gdef\SetFigFont#1#2#3{%
  \ifnum #1<17\tiny\else \ifnum #1<20\small\else
  \ifnum #1<24\normalsize\else \ifnum #1<29\large\else
  \ifnum #1<34\Large\else \ifnum #1<41\LARGE\else
     \huge\fi\fi\fi\fi\fi\fi
  \csname #3\endcsname}%
\else
\gdef\SetFigFont#1#2#3{\begingroup
  \count@#1\relax \ifnum 25<\count@\count@25\fi
  \def\x{\endgroup\@setsize\SetFigFont{#2pt}}%
  \expandafter\x
    \csname \romannumeral\the\count@ pt\expandafter\endcsname
    \csname @\romannumeral\the\count@ pt\endcsname
  \csname #3\endcsname}%
\fi
\fi\endgroup
\begin{picture}(9422,6371)(266,-6645)
\put(5581,-6046){\makebox(0,0)[lb]{\smash{\SetFigFont{9}{10.8}{rm}E}}}
\put(5356,-4111){\makebox(0,0)[rb]{\smash{\SetFigFont{9}{10.8}{rm}$\bs$}}}
\put(6871,-2701){\makebox(0,0)[rb]{\smash{\SetFigFont{9}{10.8}{rm}$\bR$}}}
\put(5596,-4921){\makebox(0,0)[rb]{\smash{\SetFigFont{9}{10.8}{rm}$\br$}}}
\put(4741,-6076){\makebox(0,0)[rb]{\smash{\SetFigFont{9}{10.8}{rm}W}}}
\end{picture}

	    $(b)$
\begin{picture}(0,0)%
\includegraphics{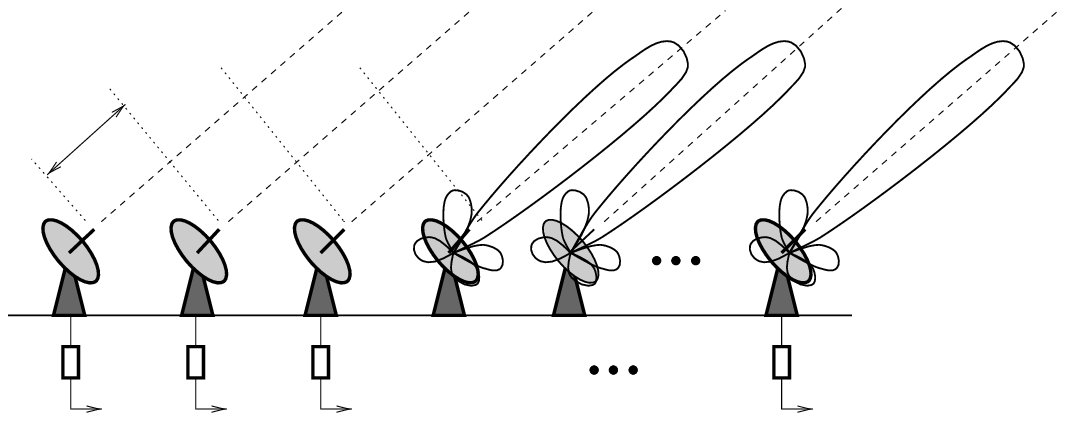}%
\end{picture}%
\setlength{\unitlength}{1973sp}%
\begingroup\makeatletter\ifx\SetFigFont\undefined
\def\x#1#2#3#4#5#6#7\relax{\def\x{#1#2#3#4#5#6}}%
\expandafter\x\fmtname xxxxxx\relax \def\y{splain}%
\ifx\x\y   
\gdef\SetFigFont#1#2#3{%
  \ifnum #1<17\tiny\else \ifnum #1<20\small\else
  \ifnum #1<24\normalsize\else \ifnum #1<29\large\else
  \ifnum #1<34\Large\else \ifnum #1<41\LARGE\else
     \huge\fi\fi\fi\fi\fi\fi
  \csname #3\endcsname}%
\else
\gdef\SetFigFont#1#2#3{\begingroup
  \count@#1\relax \ifnum 25<\count@\count@25\fi
  \def\x{\endgroup\@setsize\SetFigFont{#2pt}}%
  \expandafter\x
    \csname \romannumeral\the\count@ pt\expandafter\endcsname
    \csname @\romannumeral\the\count@ pt\endcsname
  \csname #3\endcsname}%
\fi
\fi\endgroup
\begin{picture}(10434,4002)(560,-6481)
\put(1576,-3646){\makebox(0,0)[rb]{\smash{\SetFigFont{9}{10.8}{rm}delay}}}
\put(1876,-6436){\makebox(0,0)[lb]{\smash{\SetFigFont{9}{10.8}{rm}$x_1(t)$}}}
\put(3076,-6436){\makebox(0,0)[lb]{\smash{\SetFigFont{9}{10.8}{rm}$x_2(t)$}}}
\put(4276,-6436){\makebox(0,0)[lb]{\smash{\SetFigFont{9}{10.8}{rm}$x_3(t)$}}}
\put(8701,-6436){\makebox(0,0)[lb]{\smash{\SetFigFont{9}{10.8}{rm}$x_{14}(t)$}}}
\put(1651,-6061){\makebox(0,0)[lb]{\smash{\SetFigFont{9}{10.8}{rm}$T_1$}}}
\put(2851,-6061){\makebox(0,0)[lb]{\smash{\SetFigFont{9}{10.8}{rm}$T_2$}}}
\put(4051,-6061){\makebox(0,0)[lb]{\smash{\SetFigFont{9}{10.8}{rm}$T_3$}}}
\put(8476,-6061){\makebox(0,0)[lb]{\smash{\SetFigFont{9}{10.8}{rm}$T_{14}$}}}
\put(1576,-3361){\makebox(0,0)[rb]{\smash{\SetFigFont{9}{10.8}{rm}geometric}}}
\end{picture}
  \end{center}
  \caption{$(a)$ The emitted electrical field from the celestial sphere 
  is received by a rotating telescope array; 
  $(b)$ geometrical delay compensation}
  \label{fig:space}
\end{figure}

    The signals received from the celestial sphere may be considered as
    spatially incoherent wideband random noise. It is
    possibly polarized and perhaps contains spectral absorption or
    emission lines.
    Rather than considering the emitted electric field at a location on
    the celestial sphere, astronomers try to recover the {\em
    intensity} (or brightness) $I_f(\vs)$ in the direction of
    unit-length vectors $\vs$, where $f$ is a specific frequency.
    Let $E_f(\vecr)$ be the received celestial electric field at a location 
    $\vecr$ on earth (see figure \ref{fig:space}$(a)$). Assume that the
    telescopes are identical, and let $A(\bs)$ denote the amplitude
    response of a telescope to a source in the direction $\bs$.
    The measured correlation of the electric fields 
    between two sensors $i$ and $j$ with locations $\vecr_i$ and $\vecr_j$ 
    is called a {\em visibility} and is (approximately) given by
    [\citet{perley89}]
\[
    V_f(\vecr_i,\vecr_j) 
    :=    \rE\{ E_f(\vecr_i) \overline{E_f(\vecr_j)} \}
    \;=\; \int_{\hbox{\small sky}}
	  A^2(\vs)
	  I_f(\vs) e^{-j2\pi f\; \bs^T(\vecr_i - \vecr_j)/c}\; d\Omega
    \,.
\]
    ($\rE\{\,\cdot\,\}$ is the mathematical expectation operator, 
    the superscript $^T$ denotes the transpose of a vector, and overbar denotes
    the complex conjugate.)
    Note that it is only dependent on the oriented distance $\vecr_i-\vecr_j$
    between the two telescopes; this vector is called a {\em baseline}.

    For simplification, we may sometimes
    assume that the astronomical sky is a collection of $d$
    discrete point sources (maybe unresolved). This gives
\[
    I_f(\vs) = \sum_{n=1}^{d} I_f(\vs_n) \delta(\vs - \vs_n) \,,
\]
    where $\vs_n$ is the coordinate of the $n$'th source, and thus
\begin{equation}				\label{eq:discretesource}
    V_f(\vecr_i,\vecr_j) \;=\; 
	  \sum_{n=1}^d 
	    \,
	  A^2(\vs_n)
	    I_f(\vs_n)
	    \,
	    e^{-j2\pi f\, \vs_n^T(\vecr_i - \vecr_j)/c}  \,.
\end{equation}

    Up to this point we have worked in an arbitrary coordinate system.
    For earth rotation synthesis arrays, a coordinate system is often
    introduced as follows.  We assume an array with telescopes that
    have a small field of view and that track a reference source
    direction $\bs_0$ in the sky.
    Other locations in the field of view can be written as
\[
    \bs = \bs_0 + \bsigma\,, \qquad \bs_0 \perp \bsigma \,,
\]
    (valid for small $\bsigma$) and a natural coordinate system is
\[
       \bs_0 = [0,\, 0,\, 1]^T 
       \,,
       \qquad
       \bsigma = [\ell,\, m,\, 0]^T \,.
\]
    Similarly, for a planar array, the receiver baselines can be
    parameterized as
\[
    \br_i - \br_j = \lambda [u,\, v,\, w]^T \,,
    \qquad \lambda = \displaystyle\frac{c}{f} \,.
\]
    The measurement equation in $(u,v,w)$ coordinates thus becomes
\begin{equation} 					\label{eq:fourier1}
    V_f(u,v,w)
    \;=\; e^{-j2\pi w} 	  
	  \int\!\!\!\int\; 
	  A^2(\ell,m) I_f(\ell,m)\, e^{-j 2 \pi(u\ell + v m)}\, d\ell d m
    \,.
\end{equation}
    The factor $e^{-j2\pi w}$ is caused by the {\em geometrical delay}
    associated to the reference location, and can be compensated by
    introducing a slowly time-variant delay (see figure \ref{fig:space}$(b)$).
    This synchronizes the center of the field-of-view and makes the
    reference source location appear as if it was at the north pole.
    After compensation, we arrive at a measurement equation in $(u,v)$
    coordinates only, 
\begin{equation}					\label{eq:fourier}
    V_f(u,v)
    \;=\; 
	  \int\!\!\!\int\; 
	  A^2(\ell,m) I_f(\ell,m)\, e^{-j 2 \pi(u\ell + v m)}\, d\ell d m
    \,.
\end{equation}
    It has the form of a Fourier transformation. 

    The function $V_f(u,v)$ is sampled at various coordinates $(u,v)$
    by first of all taking all possible sensor pairs $i,j$ or baselines
    $\br_i-\br_j$, and second by realizing that the sensor locations
    $\br_i$, $\br_j$ are actually time-varying since the earth
    rotates.  Given a sufficient number of samples in the $(u,v)$ domain, the
    relation can be inverted to obtain an image (the `map').

\section{ARRAY SIGNAL PROCESSING FORMULATION}		\label{sec:array}

\subsection{Obtaining the measurements}

    We will now describe the situation from an array signal processing
    point of view.  The signals received by the telescopes are
    amplified and downconverted to baseband.  A time-varying delay for
    every telescope is also introduced, to compensate for the
    geometrical delay.

    Following traditional array signal processing practices, the
    signals at this point are called $x_i(t)$ rather than $E_f(\br)$,
    and are stacked in vectors 
\[
    \bx(t) = \left[\bea{c}  x_1(t) \\ \vdots\\ x_{p}(t) \ena\right] \,,
\]
    where $p$ is the number of telescopes.  These are then processed by
    a correlation stage.

\begin{figure}
    \begin{center}
\begin{picture}(0,0)%
\includegraphics{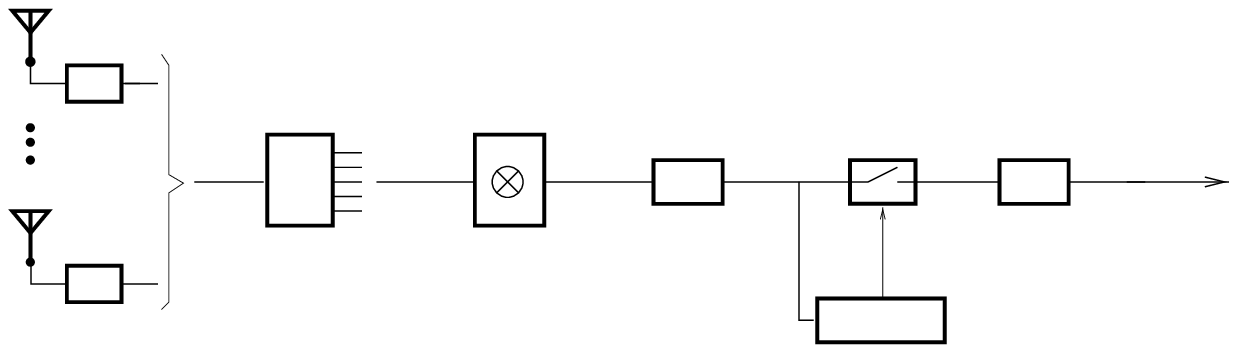}%
\end{picture}%
\setlength{\unitlength}{1533sp}%
\begingroup\makeatletter\ifx\SetFigFont\undefined
\def\x#1#2#3#4#5#6#7\relax{\def\x{#1#2#3#4#5#6}}%
\expandafter\x\fmtname xxxxxx\relax \def\y{splain}%
\ifx\x\y   
\gdef\SetFigFont#1#2#3{%
  \ifnum #1<17\tiny\else \ifnum #1<20\small\else
  \ifnum #1<24\normalsize\else \ifnum #1<29\large\else
  \ifnum #1<34\Large\else \ifnum #1<41\LARGE\else
     \huge\fi\fi\fi\fi\fi\fi
  \csname #3\endcsname}%
\else
\gdef\SetFigFont#1#2#3{\begingroup
  \count@#1\relax \ifnum 25<\count@\count@25\fi
  \def\x{\endgroup\@setsize\SetFigFont{#2pt}}%
  \expandafter\x
    \csname \romannumeral\the\count@ pt\expandafter\endcsname
    \csname @\romannumeral\the\count@ pt\endcsname
  \csname #3\endcsname}%
\fi
\fi\endgroup
\begin{picture}(16048,4805)(1750,-5550)
\put(17011,-3211){\makebox(0,0)[lb]{\smash{\SetFigFont{10}{12.0}{rm}$\mR_f^{10s}$}}}
\put(3781,-2446){\makebox(0,0)[b]{\smash{\SetFigFont{9}{10.8}{rm}$T_1$}}}
\put(3781,-4876){\makebox(0,0)[b]{\smash{\SetFigFont{9}{10.8}{rm}$T_p$}}}
\put(5806,-3211){\makebox(0,0)[rb]{\smash{\SetFigFont{10}{12.0}{rm}$\bx(t)$}}}
\put(12196,-961){\makebox(0,0)[b]{\smash{\SetFigFont{10}{12.0}{rm}10 ms }}}
\put(7246,-961){\makebox(0,0)[lb]{\smash{\SetFigFont{10}{12.0}{rm}100 kHz }}}
\put(7246,-1366){\makebox(0,0)[lb]{\smash{\SetFigFont{10}{12.0}{rm}10 $\mu$s}}}
\put(2566,-3166){\makebox(0,0)[rb]{\smash{\SetFigFont{10}{12.0}{rm}14x2}}}
\put(4681,-961){\makebox(0,0)[b]{\smash{\SetFigFont{10}{12.0}{rm}10 MHz }}}
\put(15886,-961){\makebox(0,0)[lb]{\smash{\SetFigFont{10}{12.0}{rm}10 s}}}
\put(6301,-3436){\makebox(0,0)[b]{\smash{\SetFigFont{7}{8.4}{rm}filter}}}
\put(6301,-3796){\makebox(0,0)[b]{\smash{\SetFigFont{7}{8.4}{rm}bank}}}
\put(11116,-4111){\makebox(0,0)[b]{\smash{\SetFigFont{9}{10.8}{rm}0.5--10 ms}}}
\put(13591,-5326){\makebox(0,0)[b]{\smash{\SetFigFont{9}{10.8}{rm}detector}}}
\put(15391,-3616){\makebox(0,0)[b]{\smash{\SetFigFont{9}{10.8}{rm}$\sum$}}}
\put(15391,-4111){\makebox(0,0)[b]{\smash{\SetFigFont{9}{10.8}{rm}0.5--10 s}}}
\put(8326,-3211){\makebox(0,0)[rb]{\smash{\SetFigFont{10}{12.0}{rm}$\bx_f(t_k)$}}}
\put(11116,-3616){\makebox(0,0)[b]{\smash{\SetFigFont{9}{10.8}{rm}$\sum$}}}
\put(9451,-3211){\makebox(0,0)[lb]{\smash{\SetFigFont{10}{12.0}{rm}$\bx_f\bx_f\rH$}}}
\put(12331,-3211){\makebox(0,0)[b]{\smash{\SetFigFont{10}{12.0}{rm}$\mR_f^{10ms}$}}}
\end{picture}
    \end{center}
    \caption{The astronomical correlation process}
    \label{fig:setup2}
\end{figure}

    It will be convenient to assume that $\bx(t)$ is first split by a
    bank of narrow-band sub-band filters into a collection of
    frequency-components $\bx_f(t)$.
    The main output of the telescope hardware is then a sequence of
    empirical correlation matrices $\mRh_f(t)$ of cross-correlations of 
    $\bx_f(t)$, for a set of frequencies
    $f \in \{f_k\}$ covering a 10 MHz band or so, and for a set of times
    $t \in \{t_k\}$ covering up to 12 hours.  Each correlation matrix
    $\mRh_f(t)$ is an estimate of the true covariance matrix $\mR_f(t)$,
\begin{equation}					\label{eq:avgNT}
    \mR_f(t) \;=\; {\rm E} \{ \bx_f(t) \bx_f(t)\rH \} \,,
    \qquad
    \mRh_f(t) = 
    \frac{1}{N}\sum_{n=0}^{N-1} \bx_f(t + nT) \bx_f(t + nT)\rH \,,
\end{equation}
    where the superscript $\rH$ denotes a complex conjugate transpose,
    $T$ is the sample period of $\bx_f(t)$ and $N$ is the number of
    samples over which is averaged. 
    This is drawn schematically in figure \ref{fig:setup2} (ignoring
    the detection stage for the moment).
    The matrices $\mRh_f(t)$ are stored for off-line spectral analysis and
    imaging.

    Typically, the averaging period $NT$ is in the order of 10-30~s,
    whereas each sub-band has a bandwidth in the order of 100~kHz or
    less.  Due to the sub-band filtering, the original sampling rate of
    $\bx(t)$ is reduced accordingly, resulting in $T$ in the order of
    $10\ \mu$s.

    The connection of the correlation matrices $\mR_f(t)$ to the 
    visibilities $V_f(u,v)$ in section \ref{sec:astron} is as follows.
    Each entry $r_{ij}(t)$ of the matrix $\mR_f(t)$ is a sample of this
    visibility function for a specific coordinate $(u,v)$, corresponding
    to the baseline vector $\br_i(t) - \br_j(t)$ 
    between telescopes $i$ and $j$ at time $t$:
\[
   \br_i(t) - \br_j(t) = \lambda [u_{ij}(t),\; v_{ij}(t),\; w_{ij}(t)]
\]
\[
    V_f(u_{ij}(t),v_{ij}(t)) \equiv r_{ij}(t) \,.
\]
    Note that we can obtain only a discrete set of $(u,v)$ sample points.
    Indeed, the number of instantaneous independent baselines between
    $p$ antennas is less than $\frac{1}{2}p(p-1)$. Also, using the
    earth rotation, the number of samples $\{t_k\}$ is given by the
    ratio of the observation time and the covariance averaging time
    (\eg 12 h/30 sec = 1440 samples). 

    A few remarks on practical issues are in order.
\DL
\item
    Many telescope sites including WSRT follow actually a
    different scheme where the signals are first correlated at several
    lags and subsequently Fourier transformed.  This leads to similar
    results.
\item
    The geometrical delay compensation is usually introduced only at the 
    back end of the receiver.
    At this point, also a phase correction
    is needed to compensate for the factor $e^{-j 2 \pi w_{ij}(t)}$ 
    in the measurement equation (\ref{eq:fourier1}).
    This is referred to as {\em fringe correction} \citep{thompson86}.
    Since the earth rotates, $w_{ij}(t)$ is slowly time varying, with a
    rate of change in the order of 0--10 Hz depending on source declination
    and baseline length. 
\DLE

\subsection{Matrix formulation}

    For the discrete source model, we
    can now formulate our measurement equations in terms of matrices.
    Let $\vecr_0(t_k)$ be an arbitrary and time-varying 
    reference point, typically at one of the elements
    of the array, and let us take the $(u,v,w)$ coordinates of the other
    telescopes with respect to this reference,
\[
   \br_i(t) - \br_0(t) = \lambda [u_{i0}(t),\; v_{i0}(t),\; w_{i0}(t)] \,,
   \qquad
   i = 1, \cdots, p\,.
\]
    Equation (\ref{eq:discretesource}) can then be
    written slightly different as
\[
    \bea{rcl}
    V_f(\vecr_i(t),\vecr_j(t)) &=&
	  \displaystyle\sum_{n=1}^d 
	    e^{-j2\pi f\, \vs_n^T(\vecr_i - \vecr_0)/c} 
	    \,
	    A^2(\vs_n) I_f(\vs_n)
	    \,
	    e^{j2\pi f\, \vs_n^T(\vecr_j - \vecr_0)/c}
    \\
    \Leftrightarrow
    \quad
    V_f(u_{ij}(t),v_{ij}(t)) &=& 
	  \displaystyle\sum_{n=1}^d
           
	    e^{-j 2 \pi (u_{i0}(t) \ell_n + v_{i0}(t) m_n)}
	    \,
	    A(\ell_n,m_n) 
	    \cdot \\
 & & \quad \quad I_f(\ell_n,m_n)
	    \cdot 
	    A(\ell_n,m_n) 
	    \,
	    e^{j 2 \pi (u_{j0}(t) \ell_n + v_{j0}(t) m_n)} \,.
\ena
\]
    In terms of correlation matrices, this equation can be written as 
\begin{equation}						 \label{v_mtx}
    \mR_{f}(t) \;=\; \mA_{f}(t) \mB_f \mA\rH_{f}(t)
\end{equation}
    where
\[
    \mA_{f}(t)=\left[\va_{t,f}(\ell_1,m_1),\cdots,\va_{t,f}(\ell_d,m_d)\right]
\]
and
\begin{equation}							
\label{eq:vadef} 
    \va_{t,f}(\ell,m)
    =
    \left[ \bea{c}
	e^{-j(u_{10}(t) \ell + v_{10}(t) m)} \\
	\vdots \\
	e^{-j(u_{p0}(t) \ell + v_{p0}(t) m)}
    \ena \right] A(\ell_n,m_n)
\end{equation}
\[
    \mB_f =
    \left[ \bea{ccc}
	I_f(\ell_1,m_1) & & \mbox{\bf 0}\\
	& \ddots & \\
	\mbox{\bf 0} & & I_f(\ell_d,m_d)
    \ena \right] 
\]
    The vector function $\va_{t,f}(\ell,m)$ is called
    the {\em array response vector} in array signal processing.  It
    describes the response of the telescope array
    to a source in the direction 
    $(\ell,m)$. As usual, the array response is frequency dependent.
    In this case, the response is also slowly time-varying due to the
    earth rotation.  Note, very importantly, that the function as shown
    here is completely known, since the beam shape $A(\ell,m)$ is 
    calibrated and we know the locations of the telescopes very well.

    More realistically, the array response is less perfect.
    An important effect is that each telescope may have a different complex
    receiver gain, $\gamma_i(t)$, dependent on many angle-independent
    effects such as cable losses, amplifier gains, and (slowly) varying
    atmospheric conditions.  If we take this into account, the model now becomes
\[
    \mR_{f}(t) \;=\; \mgG(t) \mA_{f}(t) \mB_f \mA\rH_{f}(t) \mgG(t)
\]
    where
\[
    \mgG(t) =
    \left[ \bea{ccc}
	\gamma_{1}(t) & & \mbox{\bf 0}\\
	 & \ddots & \\
	\mbox{\bf 0}& & \gamma_{p}(t)
    \ena \right] \,.
\]
    In future equations we will drop the dependence on $f$.

\subsection{Additive noise}

    In reality, most of the received signal consists of additive system noise.
    When this noise is zero mean, independent among
    the antennas (thus spatially white), and identically distributed, then
    it has a covariance matrix that is a multiple of the identity matrix,
    $\sigma^2 \mI$, where $\sigma^2$ is the noise power on a single antenna
    inside the subband which we consider.  Usually the noise is assumed to be 
    Gaussian. 

    The resulting model of the received covariance matrix then becomes
\begin{equation} 					\label{self_cal_model}
    \mR(t) = \mgG(t) \mA(t) \mB \mA\rH(t) \mgG(t)\rH  +  \gs^2 \mI
    \,.
\end{equation}
    Note that this assumes that the noise is introduced {\em after} the 
    varying receiver gains.
    This assumption is reasonable if the channels from the 
    low-noise amplifier (LNA) outputs to the analog-to-digital converter (ADC)
    units are equal. Otherwise, it is still reasonable to assume
    that the noise is spatially white, i.e., the noise covariance matrix is
    diagonal.  We can assume that it can be estimated using various
    calibration techniques; a simple diagonal scaling will then
    bring us back to the model (\ref{self_cal_model}).
    We further assumed that the quantization is fine, since a large
    dynamic range is needed to cope with strong interferers.

    The study of factorizations of the spatial covariance matrices such as
    shown above is the key to many array signal processing techniques.
    The knowledge of the specific structure of the array response
    vector (\ref{eq:vadef}) is of course already used in radio
    astronomy, as it enables the construction of the map using inverse Fourier
    techniques.  The main point in this paper is to demonstrate that
    also interference adds a specific structure to the covariance
    matrices.  This hopefully will allow its detection, provided our
    models are sufficiently accurate.

\section{RF INTERFERENCE}

    RF interference (RFI) usually enters the antennas through the
    sidelobes of the main beam.  It can be stronger or weaker than the
    system noise.  An important property is that it has a certain {\em
    spatial signature}, or array response vector, which becomes
    explicit in the case of narrow-band signals.

    Examples of RFI present at WSRT are television broadcasts (Smilde
    station), geolocation satellites (GPS and Glonass), taxi dispatch
    systems, airplane communication and navigation signals, wireless
    mobile communication (GSM) and satellite communication signals (Iridium).
    Thus, interferers may be continuous or intermittent, narrow-band or
    wideband, and strong or weak.  Some examples of actual interference
    are presented at the end of the section.

    Interference is usually not stationary over 10 seconds (let alone
    because of the time-varying fringe rate of 0--10~Hz), and hence it
    might be argued that it would average out from the long-term
    correlations. However, the amount of nonstationarity is often
    insufficient to provide a good and reliable protection
    \citep{thompson82}, \citep{thompson86}.

\subsection{Narrow-band interference model}

    Suppose that we have a single interferer impinging on the telescope array.
    The interfering signal reaches the array with different delays
    $\tau_i$ for each telescope.  After demodulation to baseband,
    we have
\[
    x_i(t) = a_i \, s(t-\tau_i) \, e^{- j 2 \pi  f \tau_i}\,, 
    \qquad i = 1, \cdots, p \,.
\]
    Here, $s(t)$ is the baseband signal, and
    $a_i$ represents the telescope gain in the direction of the
    interferer, including any possible attenuation of the channel. 
    Unlike much of the array signal processing literature,
    the $a_i$ are likely to be different for each telescope since the
    interferer is typically in the near field. This implies that it
    impinges on each telescope at a different angle, whereas the response of 
    the telescopes is not omni-directional.

    For narrow-band signals, time delays shorter than the inverse
    bandwidth amount to phase shifts of the baseband signal
    \citep{proakis95}.  This well-known property is fundamental to many
    phased array signal processing techniques.
    If the narrow-band assumption holds, then $s(t-\tau_i)= s(t)$ and
    the model can be simplified.

    Note that we have already assumed before that the signals are
    sub-band filtered.  Let $W$ be the bandwidth of the sub-band filters.
    In WSRT, the largest baseline is 3000~m, corresponding to a maximal
    delay of $10~\mu$s. Hence the narrow-band assumption holds if $W \ll
    100$~kHz \citet{leshem99spawc}. 
    Under this condition, we can stack the $p$ telescope
    outputs from a particular sub-band filter in a vector $\bx_f(t)$ and write
\[
    \bx_f(t) = 
    \left[\bea{c} 
	 a_1 e^{-j 2 \pi f \tau_1} \\ 
	 \vdots \\
	 a_p e^{-j 2 \pi f \tau_p}
    \ena\right] s(t) \;=:\; \ba s(t) \,.
\]
    As before, $\ba$ is an array response vector. Unlike before,
    it is not a simple or known function of the direction of the
    interferer, since we are in the near field and the sidelobes of the
    array are not calibrated.
    The vector is also called the {\em spatial signature} of the
    interfering source.
    It is slowly time varying, and we write $\ba = \ba(t)$.

    Similarly, with $q$ interferers, 
\[
    \bx_f(t) = \sum_{j=1}^q \ba_j(t) s_j(t)   
    = \mA_s(t)  \bs(t) 
    \,,
    \qquad
    \bs(t) = \left[\bea{c} s_1(t) \\ \vdots \\ s_q(t)\ena\right] \,,
    \qquad
    \mA_s(t) = \left[ \ba_1(t) \,, \cdots, \, \ba_q(t)\right] \,.
\]
    The subscript `$s$' is used to distinguish $\mA_s(t)$ from the array
    response matrix of the astronomical sources.

    The corresponding correlation matrix and its empirical estimate are
\[
    \mR(t) \;=\; {\rm E} \{ \bx_f(t) \bx_f(t)\rH \}
      \;=\; \mA_{s}(t) \mR_{s}(t) \mA_{s}\rH(t) 
    \,,
    \qquad
    \mRh(t) = \frac{1}{M} \sum_{m=0}^{M-1} \bx_f(t + m T)\bx_f\rH(t + m T)
    \,,
\]
    where $\mRh(t)$ is estimated by averaging over $M$ samples.
    The $q\times q$-matrix $\mR_{s}(t) = \rE\{\bs(t)\bs(t)\rH\}$ 
    depends on the correlation properties of the
    interfering signals. For independent interferers, it will be a diagonal
    matrix, with the $q$ interfering powers on the diagonal.

    How well the estimate fits to $\mR(t)$ depends on the stationarity of
    the scenario over the averaging interval $MT$, and is open to discussion.
    The power of television signals will be
    stationary over long periods (order tens of seconds or better). 
    At the other extreme, communication signals such as used by the GSM mobile
    communication system are time
    slotted: time is partitioned into frames of about 5~ms and frames are
    partitioned into 8 slots.  In this so-called time-division multiple
    access scheme (TDMA), 
    each user can transmit only during its slot of 0.577~ms and then
    has to be silent for 7 times this period before transmitting again
    in the next frame.  Thus, there is a short-term stationarity (over
    0.577~ms), and a cyclostationarity with periods of about 5~ms.

    The stationarity of the columns of $\mA_s(t)$ depends on the stationarity
    of the location of the interferer, its distance, the fringe rate and the 
    orientation of
    the telescopes.  With multipath propagation, a mobile interferer only
    has to move about 30~cm to create a different $\ba$-vector, 
    giving a stationarity in the order of 10--100~ms for a GSM user.
    Even for a fixed interferer such as a television station, the slow
    rotation of the telescopes as they track the sky 
    will change the $\ba$-vector within a
    fraction of a second, either because of multipath fading or because the
    interferer moves through the highly variable sidelobe pattern.
    Another source of non-stationarity is the fringe correction
    introduced at the first IF stage to compensate for the geometrical
    delay.  As the telescopes rotate, this introduces a time-varying
    phase, different to each telescope, with a rate in the range of
    $0-10$~Hz.

    The conclusion is that, for interference detection,
    $\mRh(t)$ is a useful estimate only over short
    averaging periods over which the interference is stationary, say $MT$
    in the order of milliseconds.
    Thus, $M \ll N$, where $NT \approx 10$~s, as in (\ref{eq:avgNT}).

\subsection{Overall model: astronomical signals with interference and noise}

    In summary, the model that we have derived is as follows:
\[
    \mR(t) \;=\; \mgG(t) \mA(t) \mB \mA\rH(t) \mgG(t)
	\;+\; \mA_{s}(t) \mR_{s}(t) \mA_{s}\rH(t)  
	\;+\; \sigma^2 \mI
    \,.
\]
    $\mR(t)$ is a $p\times p$ covariance matrix of which we have computed
    estimates at discrete times $t$.
    $\mA:~p\times d$ is the array response matrix of the $d$ discrete
    sources in the sky.  Its columns are known functions of the (unknown)
    locations of the sources.  It is a very wide matrix: $d \gg p$,
    and assumed stationary over 10~s.
    $\mB:~d\times d$ is a diagonal matrix (positive real)
    containing the brightness of each 
    source, and assumed time-invariant over the 
    complete observation.
    $\mgG$ are diagonal matrices (positive real) representing unknown
    and slowly varying antenna gains.

    $\mA_s:~p\times q$ is the array response matrix of the $q$ interferers.
    It is likely to be unstructured.
    We will consider cases where $q < p$, so that $\mA_s$ is tall.
    $\mR_s:~q\times q$ is the interference correlation matrix. $\mA_s$
    and $\mR_s$ are usually stationary only over very short time spans
    (order 10~ms).

    $\sigma^2 \mI$ is the noise covariance matrix, assuming white
    independent and identically distributed noise for simplicity. The
    noise power $\sigma^2$ is often rather well known.

    $\|\mA \mB \mA\rH\|$, \ie the observed power of the astronomical 
    sources, is at least two orders of magnitudes smaller than $\sigma^2$,
    and for the purpose of detection, it can be ignored.
    In contrast,
    $\|\mA_{s} \mR_{s} \mA_{s}\rH\|$ can be of comparable magnitude.

\subsection{Examples of interfering signals}
    To demonstrate a few of its features, we present some measured observations
    of RFI.\footnote{The data has been collected using the NOEMI recording
system described later in section \ref{experiment}.}

    As mentioned before, interference may be continuous or intermittent.
    A prime example of continuously present interference are television
    broadcasts.  Figure \ref{fig:tvlingen} shows a spectrogram centered at
    780.75~MHz of the German television transmitter TV Lingen, located at about
    80~km southeast of the WSRT.  The two strong peaks in the spectrum are
    the two sound carrier waves.  The received power of the TV station is much
    stronger than the WSRT system noise level, as can be seen from the fact
    that the baseband filter shape is barely visible.

    Figure \ref{fig:gsmup} shows the GSM uplink band, which contains the
    communication of mobile handsets to the base stations.  The short white
    dashes indicate the presence of (weak) interference.
    At least three channels can be seen at 902.4, 904.4, and 907.2~MHz,
    although there probably are more active channels at a lower power
    level.  The TDMA time frame format of about 5~ms consisting 8 user slots of
    0.577~ms can be recognized.  Also visible is
    a weak continuous transmission at 902~MHz. This is likely to be an 
    interference from the control building or an inter-modulation product.

    An example of the GSM downlink band (base station to mobiles)
    is shown in figure \ref{fig:gsmdown}.
    Most of the signals are continuous in time, except for a few channels
    at e.g., 942.0, 942.8, and 949.8~MHz which are time slotted. 

    Another interferer which one would like to remove from the observed data
    is the Iridium transmissions.
    Figure \ref{fig:iridium} shows a transmission of the Iridium
    satellite communication system at 1624~MHz (satellite-satellite
    communication and/or downlinks).
    It is clearly intermittent as well.

\begin{figure}
\begin{center}
 \mbox{\psfig{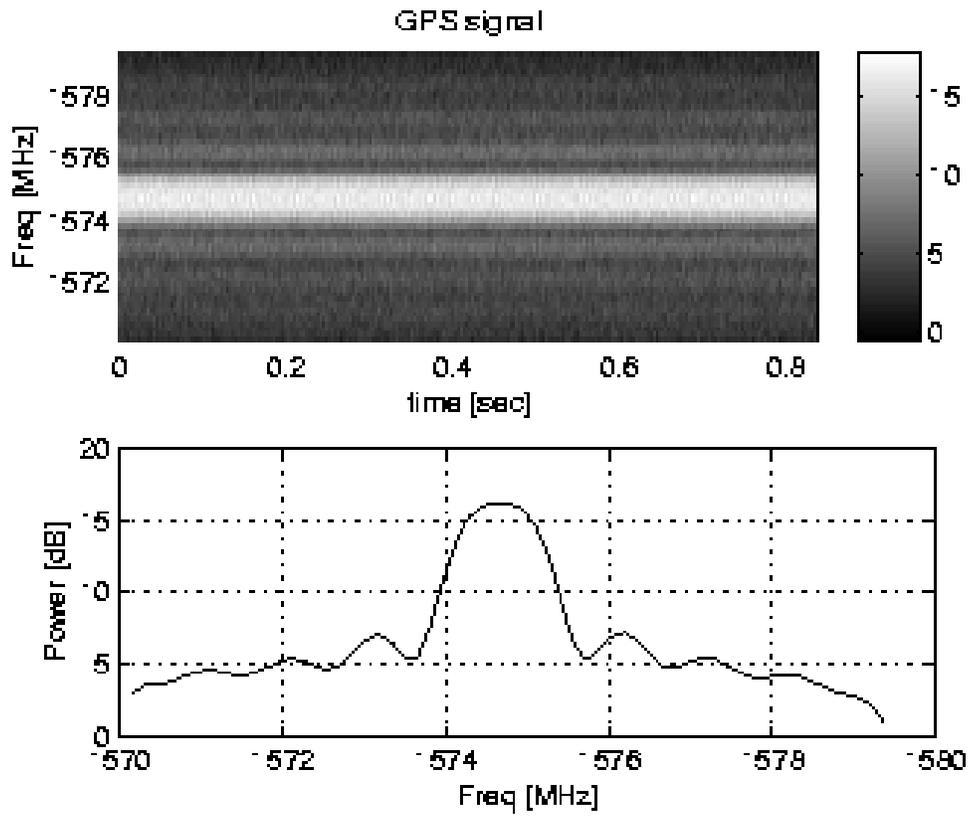}}
\end{center}
 \caption{GPS transmission, showing the civil code (BW$=1.023$ MHZ) 
superimposed on the wideband military code (BW$=10.23$ MHZ). $f_c=1575$ MHz.}
 \label{fig:GPS}
\end{figure}

    Finally a wideband interferer is the GPS satellite
    navigation signal. This is a 
    spread spectrum signal occupying a band of $10.23$~MHz. Figure 
    \ref{fig:GPS} shows a spectrogram of the GPS signal around $1575$~MHz.
    One can clearly see the superposition of the narrow ($1.023$~MHz) commonly
    available C/A signal on the wideband ($10.23$~MHz) military P-code
    signal, resulting in the peak at the center frequencies.

\section{INTERFERENCE DETECTION}		 \label{sec:detection}

\subsection{Introduction}

    Ideally, the output of the correlation process produces clean estimates
    of $\mA(t) \mB \mA(t)\rH$, once every 10~s or so. 
    In principle, we estimate it by
\begin{equation}					\label{eq:covest}
    \mRh^{10s}_f(t) 
    = \frac{1}{N} \sum_{n=0}^{N-1} \bx_f(t + n T)\bx_f\rH(t + n T)
\,,\qquad
    N T = 10\ \mbox{s}.
\end{equation}
    As we have seen, these estimates are corrupted by interference and
    additive system noise, and unknown antenna gains.
    The objective of interference detection and rejection schemes is to
    improve the {\em signal to interference and noise ratio} (SINR) at the
    output of the integrators, \ie at the 10 s level. Interference that is
    stationary at these time scales or longer can often be treated off-line.
    In this paper we 
    consider {\em online} interference detection and excision schemes, assuming
    stationarity at millisecond time scales or less.

    Many interference detection schemes exist.  They differ by the amount
    of knowledge that we can assume on the interfering
    signals, e.g., if we know the
    signal wave form, then the optimal detector has the form of a matched
    filter.  Extensions are possible if the waveform is known up to a few
    parameters such as amplitude, phase or frequency. However, usually the
    signal is modulated by a message and hence effectively unknown.  
    There are two classes of detection techniques: more or less
    deterministic methods that exploit known properties of the signals
    such as modulation type or certain periodicities, and those based
    on statistical models with unknown parameters, leading to Generalized
    Likelihood Ratio Tests (GLRT), a particular example of which is power
    detection.

    In principle, we can say that man-made interference is expected to be
    statistically different from the astronomical sources. Although
    this is a very attractive feature, it is not easy to use these
    properties for detection or excision, since the long averaging periods
    and the central limit theorem tend to jointly Gaussianize the interferers.
    For strong narrow-band interferers these methods are expected to 
    give improved suppression at an increased computational expense
    \citep{leshem98t1}. 

    Another distinction between interferers and astronomical signals is
    their spatial signature vectors.  Astronomical signals enter
    through the main lobe of the telescopes and have a very structured
    (parametrically known) array response (viz.\ (\ref{eq:vadef})),
    which is used for imaging.
    The interferers usually enter through the sidelobes and are in the
    near field, leading to unstructured $\ba$-vectors.  Also, their
    location relative to the array is not correlated with the motion of
    the earth. It might even remain fixed relative to the array during
    the complete observation period (e.g., TV transmitters). Since the
    array tracks a fixed region in the sky which moves as the earth
    rotates, the directional vector of the interference is typically
    time varying.

    From all possibilities, we consider here two schemes:
\begin{dashlist}
\item
    {\em Multichannel interference detection and excision.}
    The interference is detected at short time scales (ms), and contaminated
    samples are removed from the averaging process in (\ref{eq:covest}).
    This will work well if the interference is concentrated
    in frequency and time, as \eg in the GSM system.
\item
    {\em Spatial filtering.}  This more ambitious scheme is also suitable for 
    continuously present interference such as TV stations.  After
    detection, we estimate the spatial signature of the interferer and
    project out that dimension or otherwise subtract the signal coming from
    that direction.
\end{dashlist}

    For the purpose of power detection schemes, it is sufficient to look at
    (short-term) correlation matrices based on measurement data in a window
    of length $MT$, with $M T \approx 10\ \mbox{ms}$:
\[
    \mRh_k = \frac{1}{M} \sum_{m=0}^{M-1} \bx_f(t_k + m T)\bx_f\rH(t_k + m T)
\,,\qquad
    t_k = 0,\; M T,\; 2MT,\; \cdots \,.
\]
    If an interferer is detected in this analysis window, it is
    discarded, otherwise the data is accepted and the correlation
    matrix is used in the formation of a clean estimate of
    $\mRh^{10s}_f(t)$, as in figure \ref{fig:setup2}.  Obviously, many
    variations are possible, such as sliding window techniques, or
    discarding neighbors of contaminated samples as well (perhaps both
    in time and frequency).

    In this section we propose sub-band detection methods based on
    $\mRh_k$ and analyze their performance. Spatial filtering is
    discussed in section \ref{sec:spatfilt}.  Throughout the section, we
    will drop the subscript $k$ and write $\mR$ and $\mRh$ for simplicity.

\subsection{Single channel spectral detector}

    Detection theory is based on hypothesis testing. We test $\cH_0$: there
    is no interference, versus $\cH_1$: there is at least one interferer 
    in this band.  The implementation of this test depends on the model
    that we pose for the interferer. We will first discuss some
    particularly simple cases which will allow analysis.
    
    Thus let us consider the single-channel case first.
    We assume that there is at most a single interferer, where
    the interfering signal is i.i.d.\ Gaussian noise with unknown
    power $\sigma_s^2$. The background noise is white Gaussian with
    known power $\sigma^2$.

    Without interferer, the observed data samples $x_m \equiv x(t_m)$ 
    are complex
    normal ($\cC\cN$) distributed, with zero mean and variance $\sigma^2$. 
    With an interferer, this distribution is still complex normal, but with
    variance $\sigma_s^2 + \sigma^2$. Thus, we test the
    hypothesis
\begin{equation}						 
\label{eq:singchanhypothesis}
\bea{ll}
     \cH_0: & x_m  \;\sim\; \cC\cN(0,\, \sigma^2)\\
     \cH_1: & x_m \;\sim\;
         \cC\cN(0,\,\sigma_s^2 + \sigma^2)\,,
     \qquad m = 0, \cdots, M-1 \,.
\ena
\end{equation}
    We assume that we have available $M$ samples $\{x_m\}$, collected in a
    vector $\bx = [x_1\,,\cdots\,,x_M]$.

    This is a rather standard problem in detection theory (see \citep{kaydet}
    for an introduction).
    A Neyman-Pearson detector selects $\cH_1$ if the likelihood ratio,
\[
    L(\bx) = \frac{p(\bx; \cH_1)}{p(\bx; \cH_0)} \,,
\]
    exceeds a threshold, where $p(\bx;\cH)$ denotes the probability density
    function of $\bx$ under the hypothesis $\cH$.
    It is known that this leads to an optimal
    probability of detection, given a certain probability of false
    alarm (detecting an interferer when there is none).
    In our case, based on the model (\ref{eq:singchanhypothesis}),
    the Neyman-Pearson detector simplifies to comparing the
    total received power to a threshold $\gamma$, deciding $\cH_1$ if
    the test statistic
\[
    T(\bx) := \frac{1}{\sigma^2} \sum_{m=0}^{M-1} |x_m|^2 \quad > \quad
    \gamma \,.  
\]
    Under the above assumptions we can obtain closed form expressions
    for the probability of false alarm and the probability of detection. 
    For this, recall that the sum of squares of $M$ real i.i.d.\ zero-mean
    unit-variance Gaussian random variables has a chi-square ($\chi^2$) 
    distribution
    with $M$ degrees of freedom. Since we have complex samples,
    $T(\bx)$ is the sum-square of $2M$ real variables.
    Under $\cH_0$, these have a variance $\frac{1}{2}$,
    hence the probability of false alarm is given by 
\[
    P_{FA} \; := \; P\left\{  T(\bx) > \gamma \;;\;\cH_0 \right\}
       \quad=\quad Q_{\chi^2_{2M}} (2\gamma) 
\]
    where $Q_{\chi^2_{2M}}(\gamma)$ is the tail probability of a
    $\chi^2$ random variable with ${2M}$ degrees of freedom. It has a
    closed-form expression (cf.\ \citep{kaydet}):  
\[
    Q_{\chi^2_{2M}}(2\gamma) \; = \; e^{-\gamma}
    \sum_{k=0}^{M-1} \frac{\gamma^k}{k!} 
    \,.
\]
    Its inverse is known in terms of the inverse Gamma-function, and
    allows to select $\gamma$ to obtain a desired level of false
    alarm.
    Similarly, the probability of detection of an interference at this
    threshold $\gamma$ is given by 
\begin{equation}
    \bea{lcl}
       P_D &:=& P\{  T(\bx) > \gamma \;;\;\cH_1\} \\ 
           &=& P\{  \frac{1}{\sigma^2}\sum_{m=1}^M| x_m|^2 > \gamma 
	      \;;\;\cH_1\} \\ 
	   &=& \displaystyle P\left\{\frac{2}{\sigma^2+\sigma_s^2}
	    \sum_{m=1}^M| x_m|^2 > \frac{2\gamma}{1+\sigma_s^2/\sigma^2}
	    \;;\; \cH_1 \right\} \\ 
	   &=&
	Q_{\chi^2_{2M}}(\frac{2\gamma}{1+\INR}) 
    \ena 
\end{equation}
    where $\INR=\frac{\sigma_s^2}{\sigma^2}$ is the interference-to-noise 
ratio.

\subsection{Multichannel detector with known spatial signature}

\begin{figure}
  \begin{center}
           \mbox{ \psfig{figure=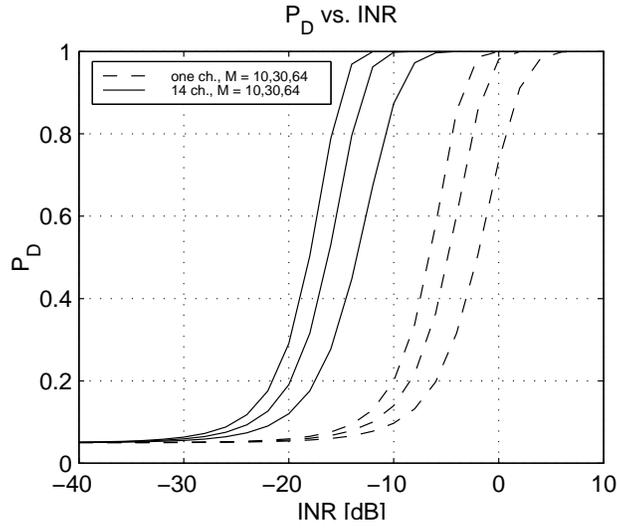,width=0.5\textwidth}}
  \end{center}
  \caption{$P_D$ vs.\ INR, for $M=10,\; 30,\; 64$, and false alarm rate
  $P_{FA} = 5\%$}
  \label{PDIa}
\end{figure}

    A significant performance improvement is possible with a
    multichannel detector. To illustrate this,
    we assume again the simple case with at most
    a single narrow-band Gaussian interferer, with
    {\em known} spatial signature vector $\ba$ in white Gaussian noise.
    The source power of the interference is denoted by $\sigma_s^2$;
    to normalize the receiver gain we set $\|\ba\|^2 := \ba\rH \ba
    = p$, where $p$ is the number of channels.
    Without interference, the data vectors $\bx_m$ are complex normal
    distributed with zero mean and covariance matrix $\sigma^2 \mI$.
    With a single interferer, the covariance matrix becomes
    $\mR = \rE\{\bx_m\bx_m\rH\} = \sigma_s^2 \ba\ba\rH + \sigma^2 \mI$.
    Thus,
\[
    \bea{ll}
       \cH_0: & \bx_m \sim \cC\cN(0, \sigma^2 \mI) \\
       \cH_1: & \bx_m \sim 
		   \cC\cN(0,\sigma_s^2 \ba\ba\rH + \sigma^2 \mI)\,,
    \qquad m = 0, \cdots, M-1 \,.
\ena
\]
    The Neyman-Pearson detector based on the data matrix $\bX =
    [\bx_1\,,\cdots,\, \bx_M]$ considers the
    estimated data covariance matrix
\[
    \mRh = \frac{1}{M} \sum_{m=0}^{M-1} \bx_m\bx_m\rH
\]
    and is given by \citep{kaydet}
\[
    T(\mX) \quad :=\quad  
	\frac{1}{\sigma^2/M} \, \frac{\ba\rH {\mRh} \ba}{\ba\rH \ba}
    \qquad
	\overset{\cH_1}{\underset{\cH_0}{\gtrless}}
    \quad
	\gamma \,.
\]
    This test is recognized as a matched spatial filter detector;
    essentially we compare the received energy in the direction $\ba$ of
    the interferer to $\sigma^2$.  If we define $y_m$ to be the output of the
    matched beamformer in the direction of $\bx_m$,
\[
   y_m = \frac{\ba\rH}{\|\ba\|} \bx_m
\]
    then
\[
    \bea{ll}
       \cH_0: & y_m \sim \cC\cN(0, \sigma^2 ) \\
       \cH_1: & y_m \sim 
		   \cC\cN(0,p \sigma_s^2 + \sigma^2 )\,,
    \qquad m = 0, \cdots, M-1 \,.
\ena
\]
    and it is seen that taking the same threshold as in
    the single channel case will provide the same false alarm probability 
    as before:
\[
    P_{FA} \;=\; P\left\{ T(\mX) > \gamma;\; \cH_0 \right\}
    \; =\; Q_{\chi^2_{2M}} (2\gamma) \,.
\]
    However, the probability of detection is now given by
\[
    P_D = P\left\{ T(\mX) > \gamma;\; \cH_1 \right\}
    \;=\; Q_{\chi^2_{2M}} (\frac{2\gamma}{1 + p\, \INR}) \,.
\]
    Figure \ref{PDIa} presents the probabilities of detection as a
    function of interference to noise ratio for a single and for
    $p=14$ channels.
    We have selected a threshold such that $P_{FA} = 5\%$, which means
    that without interference, we will throw away 5\% of the data.
    We can clearly see that the probability of detection is greatly
    improved by moving to the multichannel case.  The improvement is
    equal to the array gain, $10 \log(p) = 11.5$ dB.

\subsection{Single TDMA interferer with known spatial signature}

\begin{figure}
  \begin{center}
\begin{picture}(0,0)%
\includegraphics{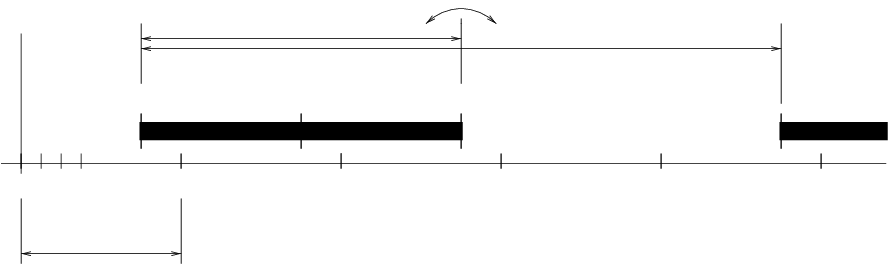}%
\end{picture}%
\setlength{\unitlength}{1263sp}%
\begingroup\makeatletter\ifx\SetFigFont\undefined
\def\x#1#2#3#4#5#6#7\relax{\def\x{#1#2#3#4#5#6}}%
\expandafter\x\fmtname xxxxxx\relax \def\y{splain}%
\ifx\x\y   
\gdef\SetFigFont#1#2#3{%
  \ifnum #1<17\tiny\else \ifnum #1<20\small\else
  \ifnum #1<24\normalsize\else \ifnum #1<29\large\else
  \ifnum #1<34\Large\else \ifnum #1<41\LARGE\else
     \huge\fi\fi\fi\fi\fi\fi
  \csname #3\endcsname}%
\else
\gdef\SetFigFont#1#2#3{\begingroup
  \count@#1\relax \ifnum 25<\count@\count@25\fi
  \def\x{\endgroup\@setsize\SetFigFont{#2pt}}%
  \expandafter\x
    \csname \romannumeral\the\count@ pt\expandafter\endcsname
    \csname @\romannumeral\the\count@ pt\endcsname
  \csname #3\endcsname}%
\fi
\fi\endgroup
\begin{picture}(13331,4911)(2089,-6022)
\put(4201,-2986){\makebox(0,0)[b]{\smash{\SetFigFont{8}{9.6}{rm}$0$}}}
\put(9001,-2911){\makebox(0,0)[b]{\smash{\SetFigFont{8}{9.6}{rm}$L$}}}
\put(4726,-2986){\makebox(0,0)[lb]{\smash{\SetFigFont{8}{9.6}{rm}Interfering slot ($\sigma_s$)}}}
\put(9001,-1411){\makebox(0,0)[b]{\smash{\SetFigFont{8}{9.6}{rm}duty cycle $\beta$}}}
\put(3601,-5971){\makebox(0,0)[b]{\smash{\SetFigFont{8}{9.6}{rm}window}}}
\put(4801,-4261){\makebox(0,0)[b]{\smash{\SetFigFont{8}{9.6}{rm}$M$}}}
\put(2401,-4261){\makebox(0,0)[b]{\smash{\SetFigFont{8}{9.6}{rm}$0$}}}
\put(3301,-4261){\makebox(0,0)[b]{\smash{\SetFigFont{8}{9.6}{rm}$m$}}}
\put(7201,-4261){\makebox(0,0)[b]{\smash{\SetFigFont{8}{9.6}{rm}$2M$}}}
\put(9601,-4261){\makebox(0,0)[b]{\smash{\SetFigFont{8}{9.6}{rm}$3M$}}}
\put(12001,-4261){\makebox(0,0)[b]{\smash{\SetFigFont{8}{9.6}{rm}$4M$}}}
\put(14401,-4261){\makebox(0,0)[b]{\smash{\SetFigFont{8}{9.6}{rm}$5M$}}}
\put(3601,-5536){\makebox(0,0)[b]{\smash{\SetFigFont{8}{9.6}{rm}analysis}}}
\end{picture}
      \hspace*{-6em}
\begin{picture}(0,0)%
\includegraphics{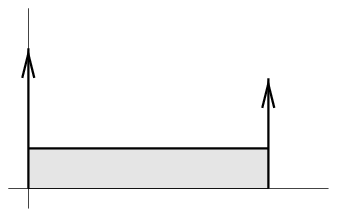}%
\end{picture}%
\setlength{\unitlength}{1263sp}%
\begingroup\makeatletter\ifx\SetFigFont\undefined
\def\x#1#2#3#4#5#6#7\relax{\def\x{#1#2#3#4#5#6}}%
\expandafter\x\fmtname xxxxxx\relax \def\y{splain}%
\ifx\x\y   
\gdef\SetFigFont#1#2#3{%
  \ifnum #1<17\tiny\else \ifnum #1<20\small\else
  \ifnum #1<24\normalsize\else \ifnum #1<29\large\else
  \ifnum #1<34\Large\else \ifnum #1<41\LARGE\else
     \huge\fi\fi\fi\fi\fi\fi
  \csname #3\endcsname}%
\else
\gdef\SetFigFont#1#2#3{\begingroup
  \count@#1\relax \ifnum 25<\count@\count@25\fi
  \def\x{\endgroup\@setsize\SetFigFont{#2pt}}%
  \expandafter\x
    \csname \romannumeral\the\count@ pt\expandafter\endcsname
    \csname @\romannumeral\the\count@ pt\endcsname
  \csname #3\endcsname}%
\fi
\fi\endgroup
\begin{picture}(11902,3720)(2437,-4495)
\put(9676,-4411){\makebox(0,0)[lb]{\smash{\SetFigFont{9}{10.8}{rm}0}}}
\put(13351,-2386){\makebox(0,0)[lb]{\smash{\SetFigFont{9}{10.8}{rm}$\frac{\alpha-1}{\alpha}\beta$}}}
\put(13201,-4411){\makebox(0,0)[b]{\smash{\SetFigFont{9}{10.8}{rm}$\sigma_s^2$}}}
\put(9376,-2011){\makebox(0,0)[rb]{\smash{\SetFigFont{9}{10.8}{rm}$1-\frac{\alpha+1}{\alpha}\beta$}}}
\put(9601,-1111){\makebox(0,0)[rb]{\smash{\SetFigFont{9}{10.8}{rm}$p(I)$}}}
\put(9376,-3436){\makebox(0,0)[rb]{\smash{\SetFigFont{9}{10.8}{rm}$\frac{1}{I_{max}}\frac{2}{\alpha}\beta$}}}
\end{picture}
  \end{center}
  \caption{(a) Interferer with slot length $L = \alpha M$ samples, power
      $\sigma_s^2$ per on-sample, and duty cycle $\beta$.
      (b)  Corresponding probability density of interference power in a
         single analysis window.  }
  \label{fig:slot}
 \label{fig:pi}
\end{figure}

    Let us now consider a TDMA signal: an
    interferer which is periodically active in
    a fraction $\beta$ of the time (see figure \ref{fig:slot}).
    Here, $0<\beta<1$ is known as
    the duty cycle of the periodic signal.  Assume that the interferer is
    present in the selected frequency band and that the duration
    of the slot in which the interferer is active is equal to $\alpha M$
    samples ${\vx}_m$, where we take $\alpha > 1$.
    Let as before $\sigma_s^2$ denote the power of a single sample of the
    interferer when it is present.

    Since the interfering slots need not be synchronized to the analysis
    window, a single interfering slot will give rise to
    two analysis windows in which the interferer is partially present, and
    possibly one or more analysis windows in which the interferer is
    present in all the samples.  Since the interferer is time-slotted with
    duty cycle $\beta$, there will also be windows that contain no
    interference.

    The corresponding probability density $p(I)$ of having a certain
    average interference power $I$ per sample in an arbitrary analysis
    window of length $M$ can be computed in closed form, as
\[
    p(I) = \left\{\bea{ll} 
	(1-\displaystyle\frac{\alpha+1}{\alpha} \beta )\,\delta(I) \,,
	    \quad& I = 0 \\[2ex]
	\displaystyle\frac{1}{I_{max}} \displaystyle\frac{2}{\alpha}\beta \,,
	    & 0 < I < I_{max} \\[2ex]
	\displaystyle\frac{\alpha-1}{\alpha} \beta \,\delta(I-I_{max}) \,,
	    \quad& I = I_{max}  \,.
    \ena\right.
\]
    It is plotted
    in figure \ref{fig:pi}, where the vertical arrows indicate the unit
    impulse function $\delta(\,\cdot\,)$.
    For example, for an interferer of strength $\sigma_s^2$
    per sample when it is on, the maximal average interference power
    per sample is obviously $\sigma_s^2$, when all samples are
    contaminated.  The probability of this is $(\alpha-1)/\alpha\,
    \beta$.  Power densities less than $\sigma_s^2$ occur with a
    uniform distribution for analysis windows that are only partly
    corrupted, at the edges of the interference slot.

    We can define
\begin{dashlist}
\item
    the average interference power per sample before detection:
\[
    I_{eff} \;=\; \int\, I\, p(I)\, dI \; = \; \beta\, \sigma_s^2 \,,
\]
\item
    the average interference power per sample after detection and blanking:
\[
    I_{res} \;=\; \int\, I\, (1-P_D(I))\, p(I)\, dI  \,,
\]
\item
    the fraction of number of samples kept after detection and blanking:
\[
    n_{res} \;=\; \int\, (1-P_D(I))\,p(I)\, dI  \,.
\]
\end{dashlist}

\begin{figure}
    \begin{center}
        \mbox{\psfig{figure=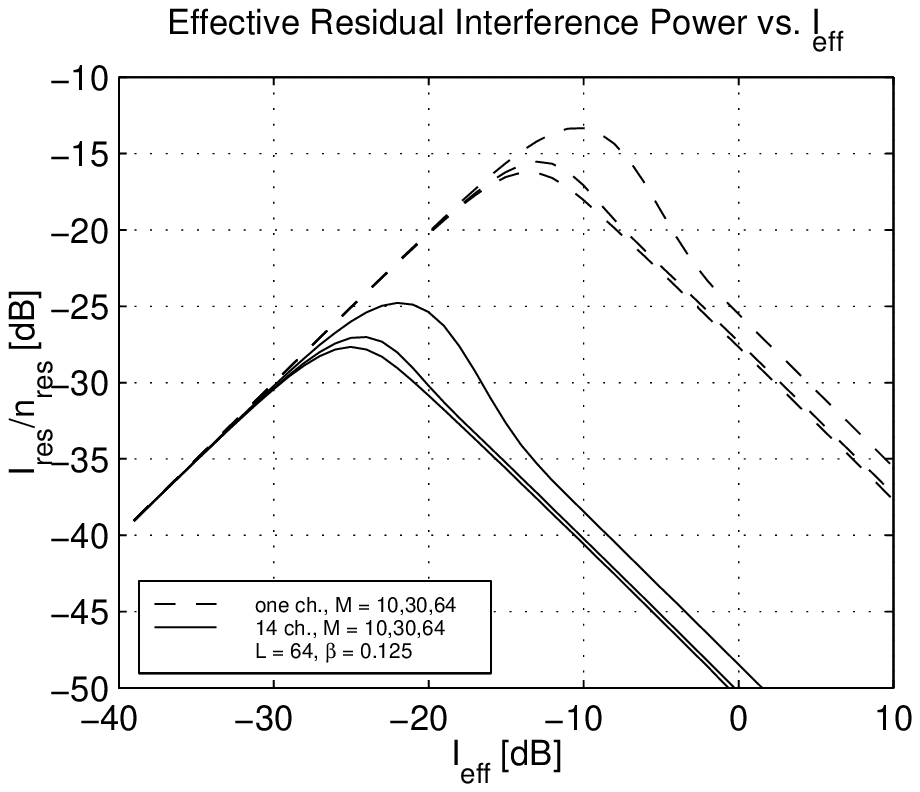,width=0.48\textwidth}}
        \mbox{\psfig{figure=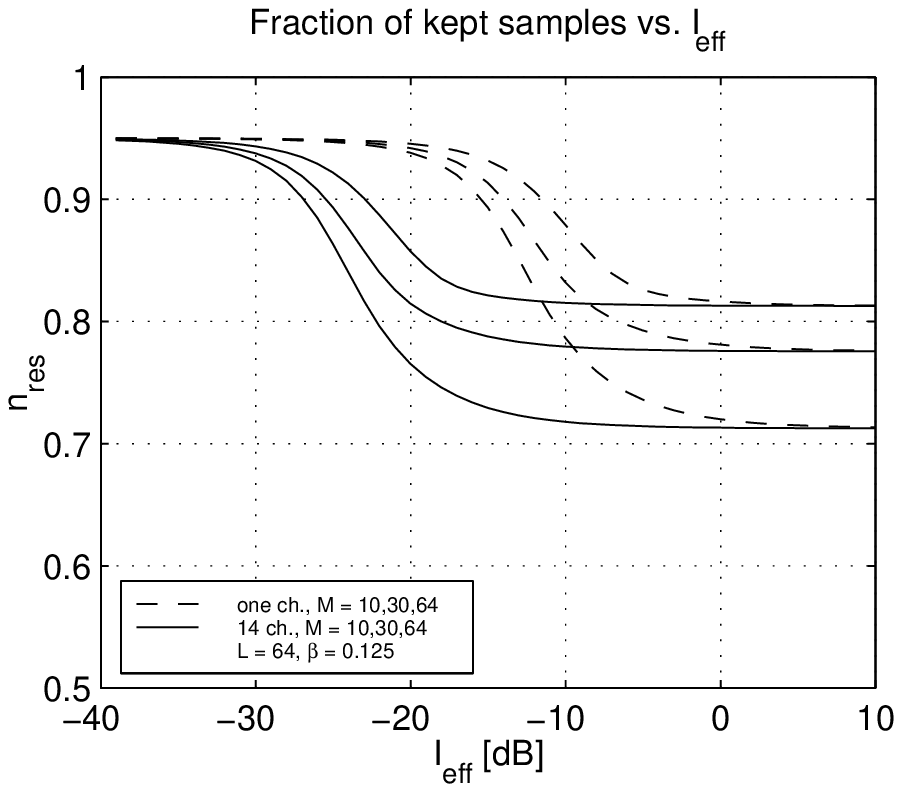,width=0.48\textwidth}}
    \end{center}
    \caption{(a) Effective residual INR after blanking
        versus effective INR at the input; (b) fraction of remaining
        samples after blanking}
    \label{PDM}
\end{figure}

    Figure \ref{PDM} shows the dependence of the residual INR as
    a function of $M$ (the number of samples in an analysis block), for an
    interferer of length $L=64$ sub-band samples, a duty cycle $\beta =
    1/8$, and a false alarm rate of $5\%$.  Obviously, very weak
    interference is not detected, and in that case we throw away $5\%$ of
    the data due to the false alarm rate.  High interference powers are
    easily detected, and almost all contaminated analysis windows will be
    detected and blanked.  Only the tails of an interfering slot might be
    missed, so that there is still some interference remaining after
    detection.  The worst case occurs for interference that is not strong
    enough to be detected all the time, but not weak enough to be
    harmless.

    Several other interesting facts can be seen in these figures.
    The most important is the large performance gain in the
    multichannel approach, as compared to a single channel.  As seen in
    figure \ref{PDIa}, the effect of using an array is to shift the
    graphs of probability of detection to the left by the array gain,
    \eg for the 14-channel detector the graph is shifted by $11.5$ dB.
    Hence, we require 11.5 dB less interference power in order to
    detect it.  However, the effective gain is given by the vertical
    distance between the graphs:  this shows the amount of interference
    suppression for a given interference power.  In figure \ref{PDM}
    the suppression can be approximately $21$ dB larger than that of
    the single antenna case.

    A second interesting phenomenon is the fact that the interference
    suppression is almost the same for a large range of analysis
    windows $M$.  Thus, we would take this window rather small, so that
    the residual number of samples is larger.  This effect is
    mainly due to the fact that the case of partial blocks with weaker
    power is less frequent as the analysis block becomes shorter.
    Further study of this model appeared in \citep{leshem99ahos}.

\subsection{Eigenvalue analysis}

    So far, we have looked at the detection problem from a rather
    idealistic viewpoint: at most 1 interferer, and a known spatial signature.
    The reason was that for this case, we could derive optimal detectors with
    closed-form expressions for the performance.
    We will now discuss an extension to more practical situations.

    Our goal is the detection of the presence of an interferer from
    observed correlation data.  As a start, let us first consider the 
    covariance matrix due to $q$ interferers and no noise,
\[
    \mR = \mA_s \mR_s \mA_s\rH
\]
    where $\mR$ has size $p\times p$, 
    $\mA_s$ has size $p\times q$ and $\mR_s$ has size $q\times q$.
    For a low number of interferers $q$, this brings us to familiar
    grounds in array signal processing, as it admits analysis by
    subspace-based techniques. We give a brief introduction here; see
    \citep{krim97} for an overview and references.

    If $q < p$, then the rank of $\mR$ is $q$ since $\mA_s$ has only $q$
    columns. Thus, we can estimate the
    number of narrow-band interferers from a rank analysis.
    This is also seen from an eigenvalue analysis: let
\[
    \mR = \mU \mLambda \mU\rH
\]
    be an eigenvalue decomposition of $\mR$, where the $p\times p$ matrix
    $\mU$ is unitary ($\mU \mU\rH = \mI$, $\mU\rH \mU = \mI$) and contains the
    eigenvectors, and the $p \times p$ diagonal matrix $\mLambda$ contains
    the corresponding eigenvalues in non increasing order ($\lambda_1 \ge
    \lambda_2 \ge \cdots \ge \lambda_p \ge 0$). Since the
    rank is $q$, there are only $q$ nonzero eigenvalues.
    We can collect these in a $q\times q$ diagonal matrix $\mLambda_s$,
    and the corresponding eigenvectors in a $p\times q$ matrix $\mU_s$, so
    that 
\begin{equation}				\label{eq:eignoisefree}
    \mR = \mU_s \mLambda_s \mU_s\rH \,.
\end{equation}
    The remaining $p-q$ eigenvectors from $\mU$ can be collected in a
    matrix $\mU_n$, and they are orthogonal to $\mU_s$ since $\mU = [\mU_s
    \;\; \mU_n]$ is unitary.  The subspace spanned by the columns of
    $\mU_s$ is called the {\em signal subspace}, the orthogonal complement 
    spanned by the columns of $\mU_n$ is known as the {\em noise subspace}
    (although this is a misnomer).
    Thus, in the noise-free case,
\[
    \mR =  \mU \mLambda \mU\rH
	= [\mU_s \quad \mU_n] 
	  \left[\bea{c|c} 
	      \mLambda_s & 0 \\ \hline 0 & 0
	  \ena\right]
	  \left[\bea{c} \mU_s\rH \\ \mU_n\rH\ena\right] 
\]
    In the presence of white noise, 
\[
    \mR \;=\; \mA_s \mR_s \mA_s\rH  + \sigma^2 \mI_p \,.
\]
    ($\mI_p$ denotes a $p\times p$ identity matrix.)
    In this case, $\mR$ is full rank: its rank is always $p$. However,
    we can still detect the number of interferers by looking at the
    eigenvalues of $\mR$. Indeed, the eigenvalue decomposition is derived
    as (expressed in terms of the previous decomposition
    (\ref{eq:eignoisefree}) and using the fact that $\mU = [\mU_s \quad \mU_n]$
    is unitary: $\mU_s \mU_s\rH + \mU_n \mU_n\rH = \mI_p$)
\begin{equation}				\label{eq:eignoisy}
    \bea{lcl}
    \mR &=& \mA_s \mR_s \mA_s\rH  + \sigma^2 \mI_p
    \\
    &=& \mU_s \mLambda_s \mU_s\rH + 
	\sigma^2 (\mU_s \mU_s\rH + \mU_n \mU_n\rH)
    \\
    &=& \mU_s (\mLambda_s + \sigma^2 \mI_q) \mU_s\rH + 
	\mU_n (\sigma^2 \mI_{p-q}) \mU_n\rH
    \\
    &=& [\mU_s \quad \mU_n] 
	  \left[\bea{c|c} 
	      \mLambda_s + \sigma^2 \mI_q & 0 \\ \hline 0 & \sigma^2 \mI_{p-q}
	  \ena\right]
	  \left[\bea{c} \mU_s\rH \\ \mU_n\rH\ena\right] 
     \\
     &=:& \mU \mLambda \mU\rH
\ena
\end{equation}
    hence $\mR$ has $p-q$ eigenvalues equal to $\sigma^2$, and $q$
    that are larger than $\sigma^2$.  {\em Thus, we can detect the number of
    interferers $q$ by comparing the eigenvalues of $\mR$ to a
    threshold defined by $\sigma^2$.}

    A physical interpretation of the eigenvalue decomposition
    can be as follows.  The eigenvectors give an orthogonal set of 
    ``directions'' (spatial signatures)\footnote{
Here, direction is not to be interpreted as the physical
direction-of-incidence of the interferer, but rather the abstract direction
of a unit-norm vector in the vector space ${\mathbb C}^{p}$.  Due to multipath,
unequal gains and fringe corrections, the physical direction-of-incidence
might not be identifiable from the spatial signature $\ba$.
}
    present in the covariance matrix, sorted in decreasing
    order of dominance.  The eigenvalues give the power of the signal
    coming from the corresponding directions, or the power of the output of
    a beamformer matched to that direction.  Indeed, let the $i$'th
    eigenvector be $\bu_i$, then this output power will be
\[
    \bu_i\rH \mR \bu_i = \lambda_i \,.
\]
    The first eigenvector, $\bu_1$, is always pointing in the direction from
    which most energy is coming. The second one, $\bu_2$, points in a
    direction orthogonal to $\bu_1$ from which most of the remaining energy
    is coming, etcetera.

    If there is no interference and only noise, then there is no dominant
    direction, and all eigenvalues are equal to the noise power.
    If there is a single interferer with power $\sigma_s^2$ and
    spatial signature $\ba$, normalized to $\|\ba\|^2 = p$, then the
    covariance matrix is $\mR = \sigma_s^2 \ba\ba\rH + \sigma^2\mI$.
    It follows from the previous that there is only one eigenvalue
    larger than $\sigma^2$.  The corresponding eigenvector is
    $\bu_1 = \ba \frac{1}{\|\ba\|}$, and is in the direction of $\ba$.
    The power coming from that direction is
\[
    \lambda_1 = \bu_1\rH \mR \bu_1 = p \sigma_s^2 + \sigma^2 \,.
\]
    Since there is only one interferer, the power coming from any other
    direction orthogonal to $\bu_1$ is $\sigma^2$, the noise power.
    Note the connection with the test statistic of the previous section,
    where we assumed that $\ba$ is known.  Since $\bu_1 = \ba
    \frac{1}{\|\ba\|}$,
\[
    \frac{ \ba\rH \mR \ba}{\ba\rH \ba} 
    = 
    \frac{ \bu_1\rH \mR \bu_1}{\bu_1\rH \bu_1}  
    = \lambda_1 \,.
\]
    Thus, the test statistic of the previous section reduces to testing the
    dominant eigenvalue of $\mR$, and knowledge of $\ba$ is in fact not
    needed. 

    With more than one interferer, this generalizes. Suppose there are two
    interferers with powers $\sigma_1$ and $\sigma_2$, and spatial
    signatures $\ba_1$ and $\ba_2$. If the spatial signatures are
    orthogonal, $\ba_1\rH \ba_2 = 0$, then $\bu_1$ will be
    in the direction of the strongest interferer, number 1 say, and
    $\lambda_1$ will be the corresponding power, $\lambda_1 = p
    \sigma_1^2 + \sigma^2$.  Similarly, $\lambda_2 = p \sigma_2^2 + \sigma^2$.

    In general, the spatial signatures are not orthogonal to each other.
    In that case, $\bu_1$ will point into the direction that is common to 
    both $\ba_1$ and $\ba_2$, and $\bu_2$ will point in the remaining
    direction orthogonal to $\bu_1$.  The power $\lambda_1$ coming from
    direction $\bu_1$ will be larger than before because it combines
    power from both interferers, whereas $\lambda_2$ will be smaller. 

\begin{figure}
\begin{center}
     \mbox{\psfig{file=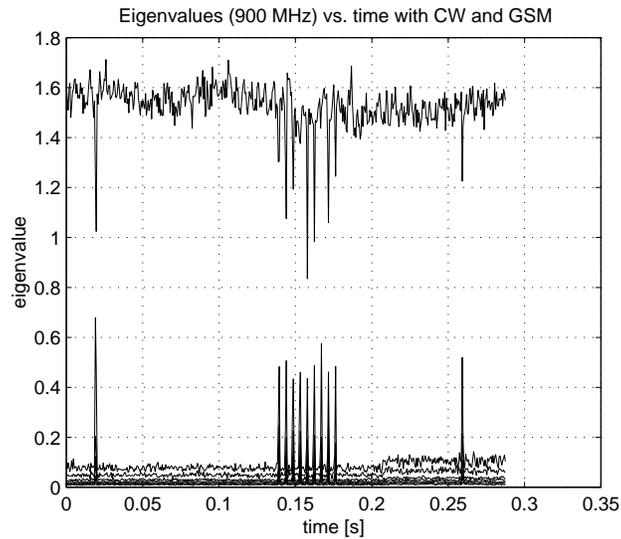,width=.5\linewidth}}
\end{center}
 \caption{Eigenstructure as a function of time}
 \label{fig:GSMeigstruct}
\end{figure}

   The covariance matrix eigenvalue structure can be nicely illustrated on
   data collected at the WSRT.  We selected a narrow band slice (52~kHz) of
   a GSM uplink data file, around
   $900$ MHz.  In this subband we have two
   interfering signals: a continuous narrow band CW signal which leaked in
   from a local oscillator, and a weak GSM signal. 
   From this data we computed a sequence of short term cross spectral
   matrices $\hat{\mR}^{0.5ms}_k$ based on $0.5$~ms averages.
   Figure \ref{fig:GSMeigstruct} shows the time evolution of the
   eigenvalues of these matrices.  The largest eigenvalue 
   is due to the CW signal and is always present.
   The GSM interference is intermittent: at time intervals where it is 
   present the number of large eigenvalues increases to two. 
   The remaining eigenvalues are at the noise floor, $\sigma^2$.
   The small step in the noise floor after $0.2$~s is due to a periodically
   switched calibration noise source at the input of the telescope front ends.

    The eigenvalue decomposition (\ref{eq:eignoisy})
    shows more than just the number of interferers.  Indeed, {\em the columns
    of $\mU_s$ span the same subspace as the columns of $\mA_s$.}
    This is clear in the noise-free case (\ref{eq:eignoisefree}), but 
    the decomposition (\ref{eq:eignoisy}) shows that the eigenvectors
    contained in $\mU_s$ and $\mU_n$ respectively
    are the same as in the noise-free case.
    Thus,
\begin{equation}					\label{eq:noisess}
    \spann(\mU_s) = \spann(\mA_s) \,, 
    \qquad \mU_n\rH \mA_s = 0 \,.
\end{equation}
    Given a correlation matrix $\mRh$ estimated from the data, we compute
    its eigenvalue decomposition. From this we can detect the rank $q$ from
    the number of eigenvalues larger than $\sigma^2$, and we can determine
    $\mU_s$ and hence the subspace spanned by the columns of $\mA_s$.
    Although we cannot directly identify each individual column of $\mA_s$,
    its subspace estimate can nonetheless be used to
    filter out the interference --- such spatial filtering algorithms are
    discussed in section \ref{sec:spatfilt}.
    Note that it is crucial that the noise is spatially white. For colored
    noise, an extension (whitening) is possible but we have to know the
    coloring.

\subsection{Multichannel detector with unknown spatial signature}

    In case we only have an estimate $\mRh$ based on a finite amount
    of samples $M$
    and the spatial signature vectors  of the interference are
    unknown, there are no optimal results. The eigenvalue analysis
    suggested that we should compare the eigenvalues to a threshold
    defined by $\sigma^2$: without interference, all eigenvalues are
    asymptotically equal to $\sigma^2$.
    We will discuss two detectors, one for the
    case where $\sigma^2$ is known, and another one for which it is
    unknown.

    If the noise power $\sigma^2$ is known, we can apply the
    (generalized) likelihood ratio test (GLRT), which leads to a method due to 
    \citet{box49} for testing the null hypothesis that $\sigma^{-2}
    \mRh = \mI$ (no interference).  The GLRT leads to a
    test statistic given by
\begin{equation}
\label{box_detect}
    T(\bX) \;:=\;
    -Mp\log \prod_{i=1}^{p} \frac{\glh_i}{\gs^2}  
\end{equation}
    where $\glh_i$ is the $i$-th eigenvalue of $\mRh$, and
    we detect an interferer if $T(\bX) > \gamma$. 
    This basically tests if all eigenvalues are equal 
    to $\sigma^2$, with a certain confidence.
    In the no-interference case, one can show that
\[
    T(\bX) \quad \sim \quad
    \chi^2_{(p+1)(p-2)}
\]
    This allows to select the  value of $\gamma$ to achieve
    a desired false alarm rate.

    If also the noise power is unknown, we propose to use
    the Minimum Description Length (MDL) detector [\citet{wax85}]. In this
    case, rather than setting a threshold based on the asymptotic
    distribution of the LRT, we try to find the
    correct model order which minimizes the description length of the
    data. The MDL rank estimator is given by
\begin{equation} 						\label{MDL}
    \hat q =\arg \min_n \MDL(n)
\end{equation}
    where
\[
    \MDL(n) \;=\;  
    -(p-n)M \log 
    \frac
	{\left(\prod_{i=n+1}^p \glh_i\right)^{\frac{1}{p-n}}}
	{\frac{1}{p-n}\sum_{i=n+1}^p \glh_i}
    \quad +\; \frac{1}{2}n(2p-n+1)\log M
\]
    and an interference is detected if $\hat q \not = 0$.  The first term
    basically tests if the geometric mean of the smallest $p-n$
    eigenvalues is equal to the
    arithmetic mean, which is only true if these eigenvalues are equal to
    each other. (The second term is a correction that grows with the number
    of unknown parameters to be estimated). Note also that the arithmetic mean 
    of the small eigenvalues is an estimate of the noise variance, so in 
    the case of testing whether $n=0$ or not the first term in the MDL 
    reduces to a sample estimate $T(\bx)$ of (\ref{box_detect}).
    This rank detector is simple to implement since it is 
    independent of the varying SINR in the system.  A disadvantage is that
    the false alarm rate is not known and not fixed.

    Finally a simple option which can be used to limit the false alarm rate
    is to collect a number of processing blocks, sort them
    according to the value of the statistic $T(\bx)$, defined in 
    (\ref{box_detect}) and throw away a given percentage with the highest 
    score. This is conceptually simple but needs more memory available.

    Experimental results on multichannel blanking are presented in section
    \ref{sec:experiments}.

\section{SPATIAL FILTERING}				\label{sec:spatfilt}

   Let us now assume that we have obtained a covariance matrix $\mR$, 
   which contains the rather weak
   covariance matrix of the astronomical sources (visibilities) $\mR_v$,
   plus white noise.
   Suppose also that there is an interferer with power $\sigma^2_s$:
\[
   \mR \;= \; \mR_v \;+\; \sigma^2_s \ba \ba\rH \; + \; \sigma^2 \mI  \,.
\]
   In the previous section, we considered schemes to detect the
   interferer from the eigenvalues of $\hat{\mR}$, a short-term estimate of
   $\mR$. After detection,
   we proposed to discard $\hat{\mR}$ from a longer-term average
   if it is found to be contaminated, 
   but what if the interferer is present all the time? In that case, it is
   more suitable to try to suppress its contribution $\sigma^2_s \ba \ba\rH$.
   This leads to {\em spatial filtering} techniques.

\subsection{Projecting out the interferer}

   An elementary form of spatial filtering 
   is to null all energy with spatial signature $\ba$. To
   this end, we can introduce the $p\times p$ projection matrix
\[
   \bP_\ba^\perp = \mI - \ba(\ba\rH \ba)^{-1} \ba\rH \,.
\]
   $\bP_\ba^\perp$ is a projection because $\bP_\ba^\perp
   \bP_\ba^\perp = \bP_\ba^\perp$. 
   It is also easily seen that $\bP_\ba^\perp \ba = {\bf 0}$: this direction is
   projected out.
   If we denote by
   $\mRt$ the filtered covariance matrix, we obtain
\begin{equation}				\label{eq:noiseproj}
      \mRt := 
      \bP^\perp_\ba \mR\bP^\perp_\ba
      \;=\; \bP^\perp_\ba \mR_v  \bP^\perp_\ba \;+\; \sigma^2 \bP^\perp_\ba 
      \,.
\end{equation}
   Thus, the interference is removed by the projection.
   At the same time, the visibility
   matrix is modified by the projections, and the noise is not white
   anymore, since one dimension is missing.  The imaging stage has to be
   aware of this, which is the topic of \citep{leshem99u2}.

   In general, $\ba$ is not known. However, note that 
   we do not need $\ba$, but only a projection matrix to project it out.
   Recall from equation (\ref{eq:eignoisy}) the
   eigenvalue decomposition of $\mR$, and in particular the matrix
   containing an orthonormal basis of the ``noise subspace'' $\mU_n$, which
   is the orthogonal complement of $\ba$, with $p-1$ columns.
   According to (\ref{eq:noisess}), $\mU_n\rH \ba = 0$. It is now
   straightforward to show that 
\begin{equation}						
\label{eq:noiseprojectdef}
   \bP_\ba^\perp = \mU_n \mU_n\rH
\end{equation}
   Indeed, since $\mU_n\rH \mU_n = \mI_{p-1}$,
\[
   \bP_\ba^\perp \bP_\ba^\perp 
      \;=\; \mU_n \mU_n\rH \mU_n \mU_n\rH
      \;=\; \mU_n \mU_n\rH 
      \;=\; \bP_\ba^\perp
\]
   and 
\[
   \bP_\ba^\perp \ba \;=\; \mU_n \mU_n\rH  \ba \;=\; {\bf 0} \,.
\]
   Thus, we can compute the required projection matrix directly from
   the eigenvalue decomposition of $\mR$.

   Expression (\ref{eq:noiseprojectdef})
   can immediately be generalized to the more general
   case of $q<p$ interferers and unknown $\ba$-vectors. 
   Indeed, in this case, the projection onto the complement of the
   $\mA_s$-matrix of the interference is given by 
\[
   \bP_{\mA_s}^\perp = \mI - \mA_s(\mA_s\rH \mA_s)^{-1} \mA_s\rH =
   \mU_n \mU_n\rH 
\]
   Note that we do not have to know $\mA_s$: the relevant noise subspace is
   estimated from the eigenvalue decomposition of $\mR$.   This hinges
   upon the fact that the noise covariance is white (in general:
   known), and the visibility matrix $\mR_v$ is insignificant at these
   time scales (otherwise, it might disturb the eigenvalue
   decomposition).

   As an alternative to
   (\ref{eq:noiseproj}), we can define another filtered covariance matrix
\begin{equation}				
\label{eq:noisedimred}
      \mRt := \mU_n\rH \mR \mU_n
      \;=\; \mU_n\rH\mR_v\mU_n  \;+\; \sigma^2 \mI_{p-q} \,,
\end{equation}
   where we have used $\mU_n \perp \mA_s$ and $\mU_n\rH \mU_n = I_{p-q}$.
   In this case, $\mRt$ has size $(p-q) \times (p-q)$.  Although smaller,
   this matrix contains the same information as $\bP^\perp_\ba
   \mR\bP^\perp_\ba$.  Besides the dimension reduction, an
   advantage of this scheme is that the noise term stays white.

\subsection{Keeping track of projections}	\label{sec:projtrack}

    Since the projections alter the visibility data in $\mR_v$, it is
    essential, for the purpose of imaging, to store the linear
    operation represented by the projections.  At the same time, it
    might be necessary to adapt the projection several times per
    second, since the $\ba$-vectors of interferers are time-varying.
    Hence, in the construction of the 10~s correlation average from
    short-term projected correlation matrices, we also
    have to construct the effective linear operation.

    Let $\mR_k$ denote the short-term correlation matrix, where 
    $k = 0, 1, \cdots, N-1$ is the time index, and $N$ is the number of
    short-term matrices used in the long-term average.
    Denote for generality the linear operation representing the projection
    by $\mL_k$, where $\mL_k
    = (\mU_n)_k (\mU_n)_k\rH$ in the first filtering scheme (equation
    (\ref{eq:noiseproj})), and $\mL_k = (\mU_n)_k\rH$ in the second (equation
    (\ref{eq:noisedimred})).
    Consider now the short-term filtered averages, 
\[
    \mRt_k := 
    \mL_k \mR_k \mL_k\rH
    \;=\; \mL_k \mR_v  \mL_k\rH \;+\; \sigma^2 \mL_k \mL_k\rH
    \,,\qquad
    k = 0, 1, \cdots, N-1
    \,.
\]
    By simply averaging these, the long-term average will be
\[
    \mRt^{10s}
    \;=\; \frac{1}{N} \sum_{k=0}^{N-1} \mRt_k
    \;=\; \frac{1}{N} \sum_{k=0}^{N-1} \mL_k \mR_k \mL_k\rH
    \,.
\]
    The $\mL_k$ appear here at both sides of $\mR_k$.  
    To move them to one
    side, we make use of the general expression
\[
    \vect(\mA \mB \mC) = (\mC\rT \otimes \mA) \vect(\mB)
\]
    where $\otimes$ denotes a Kronecker product, and $\vect(\cdot)$ the
    column-wise stacking of a matrix into a vector,
\[
    \mA \otimes \mB := 
    \left[\bea{ccc}  
	a_{11} \mB &  a_{12} \mB & \cdots \\
	a_{21} \mB &  a_{22} \mB & \cdots \\
	\vdots   &           & \ddots \ena\right]
\]
\[
    \mA = [ \ba_1 \quad \ba_2 \quad \cdots ]
    \quad\Rightarrow \quad
    \vect(\mA) :=  \left[\bea{c} \ba_1 \\ \ba_2 \\ \vdots \ena\right]
\]
    In this case, we obtain
\[
\bea{lcl}
    \vect(\mRt^{10s})
    &=& \frac{1}{N} \sum \left[(\mLb_k\otimes \mL_k)\vect(\mR_k)\right]
    \\
    &=& \left[\frac{1}{N} \sum \mLb_k\otimes \mL_k\right]
	\vect(\mR_v)
	\;+\; 
	\sigma^2 \left[\frac{1}{N} \sum \mLb_k\otimes \mL_k\right]
	\vect(\mI_p) 
    \\
    &=& \mC \vect(\mR_v) \;+\; \sigma^2 \mC \vect(\mI_p)
\ena
\]
    where 
\[
    \mC \;:=\; \frac{1}{N} \sum_{k=0}^{N-1} \mLb_k\otimes \mL_k
\]
    and the overbar denotes complex conjugation.
    $\bC$ is the effective linear mapping of entries of $\mR_v$ to entries
    of $\mRt^{10s}$.
    For the imaging step, we have to know how $\mRt^{10s}$ depends on $\mR_v$.
    Thus, we have to construct and store $\mC$ along with $\mRt^{10s}$.
    Unfortunately, it is a rather large matrix: 
    $p^2 \times p^2$ in the first filtering scheme, and $(p-q)^2\times
    p^2$ in the second.
    Another problem for imaging might be that the noise contribution on
    $\mRt^{10s}$ is no longer white, but determined by $\mC$. Two possible
    remedies are
\begin{dashlist}
\item
    Assume that the $\ba$-vectors were sufficiently variable over the time
    interval. In that case, $\mC$ is likely to be of full rank and
    thus invertible, and we can construct
\[
    \mC^{-1} \vect(\mRt^{10s}) = \vect(\mR_v) \;+\; \sigma^2  \vect(\mI_p)
    \,.
\]
    By unstacking the result, 
    we recover our interference-free model $\mR_v + \sigma^2 \mI$.
    However, the inversion of $\mC$ might be a formidable, and numerically
    dubious, task.
\item
    If we take $\mL_k = (\mU_n)_k\rH$ as in (\ref{eq:noisedimred}), then
    the noise contribution on each $\mRt_k$ is white. We can average the
    $\mRt_k$ if they have the same dimension, \ie $p-q$ where the number of
    interferers $q$ is constant over the interval. In that case,
\[
    \sigma^2 \frac{1}{N} \sum_{n=0}^{N-1} (\mU_n)_k\rH (\mU_n)_k 
    \;=\; \sigma^2 \mI_{p-q}
\]
    so that the noise contribution on $\mRt^{10s}$ is white. Note that
    no inversion is necessary.
\end{dashlist}

    If we do not invert $\mC$ then the observed visibilities
    $V(u_{ij},v_{ij})$ in the matrix $\mR_v$ are modified by some
    (known) linear combination.  This has implications for the
    synthesis imaging step.  The usual construction of an
    image using inverse Fourier transformation (based on
    (\ref{eq:fourier})) now gives rise to a point-source image
    convolved with a {\em space-varying} point spread function (``dirty
    beam'').  Since the point spread function is known at every
    location in the image, it is still possible to correct for it using an
    extension of the usual CLEAN deconvolution algorithm. Details are
    in \citep{leshem99u2}.

    Since $\mC$ is a factor $p^2$ larger than $\mRt^{10s}$, it might in
    fact be more efficient to store the sequence of spatial filters
    $\mL_k$. This is the case if $\mL_k$ is to be updated at time
    scales of $10~\mbox{s}/p^2 = 50$~ms or longer.

\subsection{Other spatial filtering possibilities}

    Without going into too much detail, we mention a few other
    possibilities for spatial filtering and interference cancellation.
    Suppose there is a single interferer,
\[
    \mR = \mR_v + \sigma_s^2 \ba\ba\rH + \sigma^2\mI \,.
\]

\begin{figure}
\begin{center}
\begin{picture}(0,0)%
\includegraphics{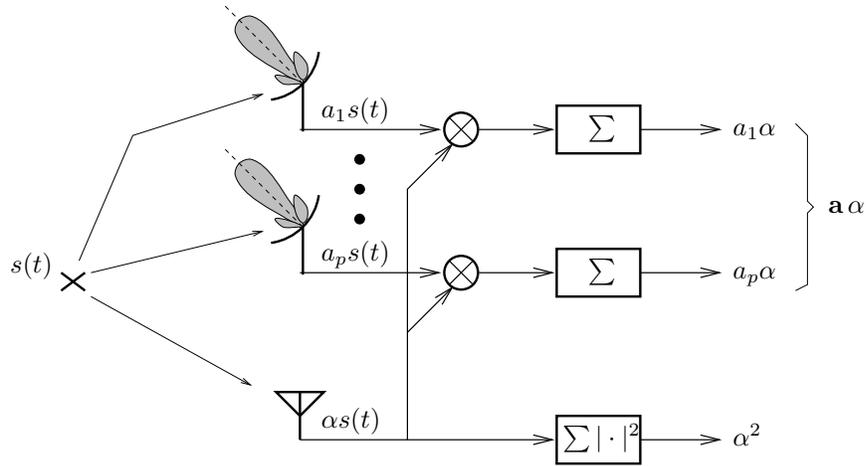}%
\end{picture}%
\setlength{\unitlength}{1973sp}%
\begingroup\makeatletter\ifx\SetFigFont\undefined
\def\x#1#2#3#4#5#6#7\relax{\def\x{#1#2#3#4#5#6}}%
\expandafter\x\fmtname xxxxxx\relax \def\y{splain}%
\ifx\x\y   
\gdef\SetFigFont#1#2#3{%
  \ifnum #1<17\tiny\else \ifnum #1<20\small\else
  \ifnum #1<24\normalsize\else \ifnum #1<29\large\else
  \ifnum #1<34\Large\else \ifnum #1<41\LARGE\else
     \huge\fi\fi\fi\fi\fi\fi
  \csname #3\endcsname}%
\else
\gdef\SetFigFont#1#2#3{\begingroup
  \count@#1\relax \ifnum 25<\count@\count@25\fi
  \def\x{\endgroup\@setsize\SetFigFont{#2pt}}%
  \expandafter\x
    \csname \romannumeral\the\count@ pt\expandafter\endcsname
    \csname @\romannumeral\the\count@ pt\endcsname
  \csname #3\endcsname}%
\fi
\fi\endgroup
\begin{picture}(10443,5820)(433,-6094)
\put(8026,-3736){\makebox(0,0)[b]{\smash{\SetFigFont{10}{12.0}{rm}$\sum$}}}
\put(8026,-1936){\makebox(0,0)[b]{\smash{\SetFigFont{10}{12.0}{rm}$\sum$}}}
\put(4501,-3511){\makebox(0,0)[lb]{\smash{\SetFigFont{10}{12.0}{rm}$a_ps(t)$}}}
\put(4501,-1711){\makebox(0,0)[lb]{\smash{\SetFigFont{10}{12.0}{rm}$a_1s(t)$}}}
\put(4501,-5611){\makebox(0,0)[lb]{\smash{\SetFigFont{10}{12.0}{rm}$\alpha s(t)$}}}
\put(9676,-3736){\makebox(0,0)[lb]{\smash{\SetFigFont{10}{12.0}{rm}$a_p\alpha$}}}
\put(9676,-1936){\makebox(0,0)[lb]{\smash{\SetFigFont{10}{12.0}{rm}$a_1 \alpha$}}}
\put(10876,-2911){\makebox(0,0)[lb]{\smash{\SetFigFont{10}{12.0}{rm}$\ba\,\alpha$}}}
\put(9676,-5836){\makebox(0,0)[lb]{\smash{\SetFigFont{10}{12.0}{rm}$\alpha^2$}}}
\put(8026,-5836){\makebox(0,0)[b]{\smash{\SetFigFont{10}{12.0}{rm}$\sum|\cdot|^2$}}}
\put(1126,-3661){\makebox(0,0)[rb]{\smash{\SetFigFont{10}{12.0}{rm}$s(t)$}}}
\end{picture}
\end{center}
    \caption{Estimation of $\ba$ using a reference antenna}
    \label{fig:refant}
\end{figure}

\begin{dashlist}
\item
   {\em Subtraction.}  With an estimate of $\ba$ and $\sigma_s^2$,
   we can try to subtract it from the covariance data:
\begin{equation}				\label{eq:intsubtract}
      \mRt = \mR - \hat{\sigma}_s^2 \bah\bah\rH  \,.
\end{equation}
   Without other knowledge, the best estimate of $\ba$ is the dominant
   eigenvector, $\bu_1$, of $\mR$, and likewise
   the best estimate of $\sigma_s^2$ is $\lambda_1 - \sigma^2$. 
   Since both of these are derived from $\mR$, it turns out to be not too
   different from the projection scheme. Indeed, if we look at
\[
    (\mI - \alpha \bu_1\bu_1\rH) \mR (\mI - \alpha \bu_1\bu_1\rH)
    \;=\; \mR - \bu_1 \bu_1\rH \lambda_1(2 \alpha - \alpha^2)
\]
   we can make it equal to (\ref{eq:intsubtract}) by
   selecting $\alpha$ such that $\lambda_1(2 \alpha - \alpha^2) =
   \hat{\sigma}_s^2$.   The projection scheme had $\alpha = 1$.

   Our point here is that also subtraction can be represented by a
   two-sided linear operation on the correlation matrix. 
   Consequently, the visibility matrix $\mR_v$ is altered,
   and hence the corrections mentioned in
   section \ref{sec:projtrack} are in order.
\item
    {\em Subtraction of a reference signal.}
    If we have a reference antenna that receives a `clean' copy of the
    interfering signal, then we might try to subtract this reference signal
    from the telescope signals.  There are many adaptive schemes for doing
    so, \eg the LMS algorithm \cite{haykin95}.
    The general scheme is as illustrated in figure
    \ref{fig:refant}. 
    In this figure, the $\ba$-vector of the interferer is found by 
    cross-correlating with the reference antenna. We also estimate its power.
    After correcting for the noise power on the reference antenna, we can 
    reconstruct and subtract $\ba s(t)$.

    This scheme is rather similar to the original projection approach where we
    reduce the dimension to the noise subspace, 
    \viz equation (\ref{eq:noisedimred}).
    The main difference is
    that, now, we reduce the dimension from $p+1$ antennas back to $p$ 
    antennas, so there is no loss of dimensions from the astronomy point of
    view.
    It appears that this 
    only has advantages if the reference antenna has a better INR 
    than the telescopes.  Also, we need as many reference antennas as there
    are interferers.
\end{dashlist}

    As with the projection technique, all of these forms of spatial
    filtering modify the observed visibilities in the matrix $\mR_v$ by
    a known linear combination, with implications for the
    synthesis imaging step \citep{leshem99u2}.

\section{MULTICHANNEL BLANKING: EXPERIMENTAL RESULTS}\label{sec:experiments}
\label{experiment}

    To test the blanking and filtering algorithms, we have attached the
    WSRT antennas to a multi-channel data recorder that can collect a 
    few seconds of data at 20~MHz rate and store it on CDROM.
    This enabled us to record a variety of actual interference and process
    it off-line.
    In this section, we demonstrate the performance of the blanking
    algorithm by adding GSM observations to ``clean'' galactic 3C48
    data, in a variety of scalings.  The results are quite good, as it
    is possible to recover a 3C48 absorption line which was completely
    masked by the GSM interference.

\subsection{Experimental setup}

    The data recorder has been acquired in the context of the STW NOEMI
    project, a cooperation between Delft University of Technology and 
    ASTRON/NFRA. 
    It basically consists of an industrial PC with four PCI.212 sampling boards.
    Each board contains two ADCs, and the boards are synchronized
    so that in total eight telescope channels can be sampled simultaneously. 
    The ADCs have a resolution of 12 bit
    with sampling rates of 20~MHz down to 0.313~MHz in steps of a factor
    of 2. After collecting a batch of data, it can be copied into system
    memory (384~MB), previewed and stored onto CDROM.\footnote{
We would like to thank G.~Schoonderbeek for programming the data
acquisition software.}

\begin{figure}
\begin{center}
   \mbox{\psfig{figure=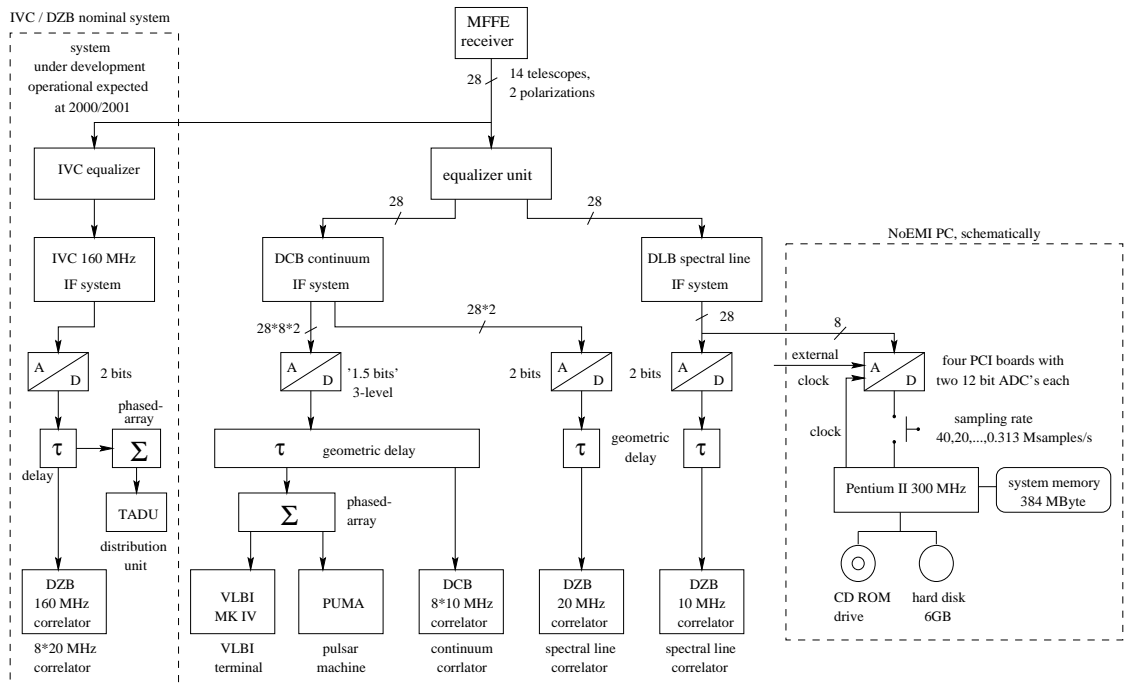,width=0.9\textwidth}}
   \caption{Overview of main WSRT systems with the NOEMI data recorders}
   \label{fig:wsrtscheme}
\end{center}
\end{figure}

    Fig.~\ref{fig:wsrtscheme} shows an overview of the WSRT
    system to indicate the point where the NOEMI data recorder was connected.
    The WSRT is an East-West linear array of fourteen
    telescope dishes, mostly spaced at 144~m.
    Each dish is equipped with a front-end receiver that can be tuned to
    several frequency bands. Both polarizations (X and Y) are received.
    The resulting $14\times 2$ channels are amplified,
    filtered, down-converted to an intermediate frequency (IF) range
    around 100~MHz, and transported to the main building via coaxial
    cables. Here, the signals are fed to the equalizer unit which
    compensates for the frequency dependent attenuation in the
    ground cables. The equalizer unit has outputs for the broadband
    continuum system (DCB, 8 bands of 10~MHz) and for the spectral line
    system (DLB, 10~MHz). In the equalizer unit and in the DCB/DLB IF
    systems are mixers, amplifiers and filter units which take care of
    the baseband conversion and filtering. At baseband the signals are
    digitized to 2-bit resolution, a correction is applied for the
    geometrical delay differences between the telescopes, and the signals
    are correlated (in pairs) in the DZB/DCB correlators.
    The NOEMI recorder is connected at the output of the DLB
    spectral line IF system. Of the $14\times 2$ available telescope
    channels, a selection of eight are connected to the NOEMI ADC samplers.
    The geometrical delay compensations and fringe rate corrections were not
    included in the measurements.\footnote{
       Since these fringe rates are in the order of 0--10~Hz, this has no
       consequences for the detection of interference based on 
       short-term correlation matrices, with typical integration periods
       in the order of milliseconds.
}

    The WSRT system contains also calibration noise sources, which are
    switched on for a 1.25s period every 10 seconds.  For regular WSRT
    observations these noise sources are used for system noise and gain
    calibration purposes. In some of the observed NOEMI data sets these
    noise sources are clearly visible as a 5--15\% power step.

    Two important tests have been applied to the recording system. The
    synchronization of the channels was checked by applying a common
    wideband signal and was found to be in order. The cross-talk
    between the channels was measured by applying a signal to only one
    of the channels and looking at the leakage into the other
    channels.  The power insulation between two channels on the same
    PCI board is found to be 51~dB (0.28\% in voltage), and at least
    90~dB (0.0032\% in voltage) across boards. This is sufficient for
    spectral line work and for RFI mitigation tests.

\subsection{Clean 3C48 absorption data}

\begin{figure}
\begin{center}
    \mbox{\psfig{figure=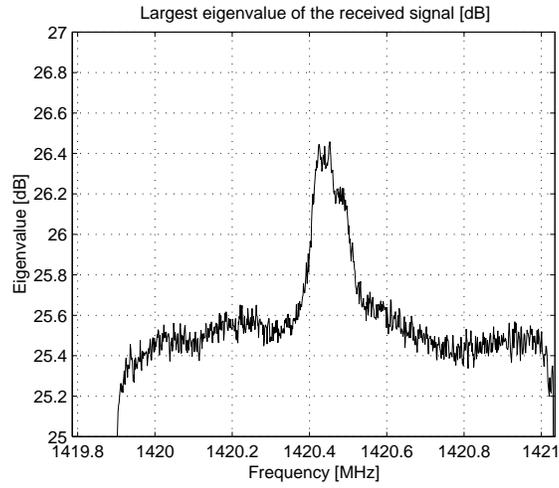,width=0.45\textwidth}}
\end{center}
    \caption{3C48: largest eigenvalue of the covariance matrix}
    \label{fig:PSD14}
\end{figure}

    To compare our off-line frequency domain correlation process 
    based on recorded data
    to the online Westerbork correlator (the DZB)
    we have made an interference-free observation of the galactic
    HI absorption of 3C48, a spectral line at 1420~MHz.  Figure
    \ref{fig:PSD14} shows the estimate of the power spectral density of
    the received signal based on the largest eigenvalue of the
    covariance matrix.

    The coherency (correlation coefficient) 
    of signals $x_i$ and $x_j$ at the output of telescopes
    $i$ and $j$ is defined as 
\begin{equation}					\label{eq:coherencydef}
    \rho_{ij}(f) \;=\; 
	\frac{\rE(x_i(f) \bar{x}_j(f))}
	     {\sqrt{\rE(|x_i(f)|^2)\rE(|x_j(f)|^2)}} 
    \;=\; 
    \frac{R_{ij}(f)}{\sqrt{R_{ii}(f) R_{jj}(f)}}
    \,.
\end{equation}
    Since all telescopes are tracking the same source $s$, we have that
    $x_i = \alpha_i s + n_i$ where $n_i$ is the noise at telescope $i$.  With
    uncorrelated noise of power $\rE(|n_i|^2) = \sigma^2$, and a source
    power of $\sigma_s^2$, it follows that 
\[
    \rho_{ij}(f) = 
	\alpha_i\bar{\alpha}_j 
	\frac{\sigma_s^2(f)}{\sigma_s^2(f) + \sigma^2(f)} 
    \qquad (i \neq j)
    \,.
\]
    Thus, the theoretical value of the coherency is constant over all
    nonzero baselines, and can be estimated based on the parameters of
    3C48 and knowledge of the receiver gains and noises.  
    These theoretically expected (asymptotic) values can then be compared
    to the computed coherencies of the recorded observation using
    (\ref{eq:coherencydef}), and can also be compared to
    the coherency measured with the DZB hardware.

\begin{figure}
    $(a)$
    \mbox{\psfig{figure=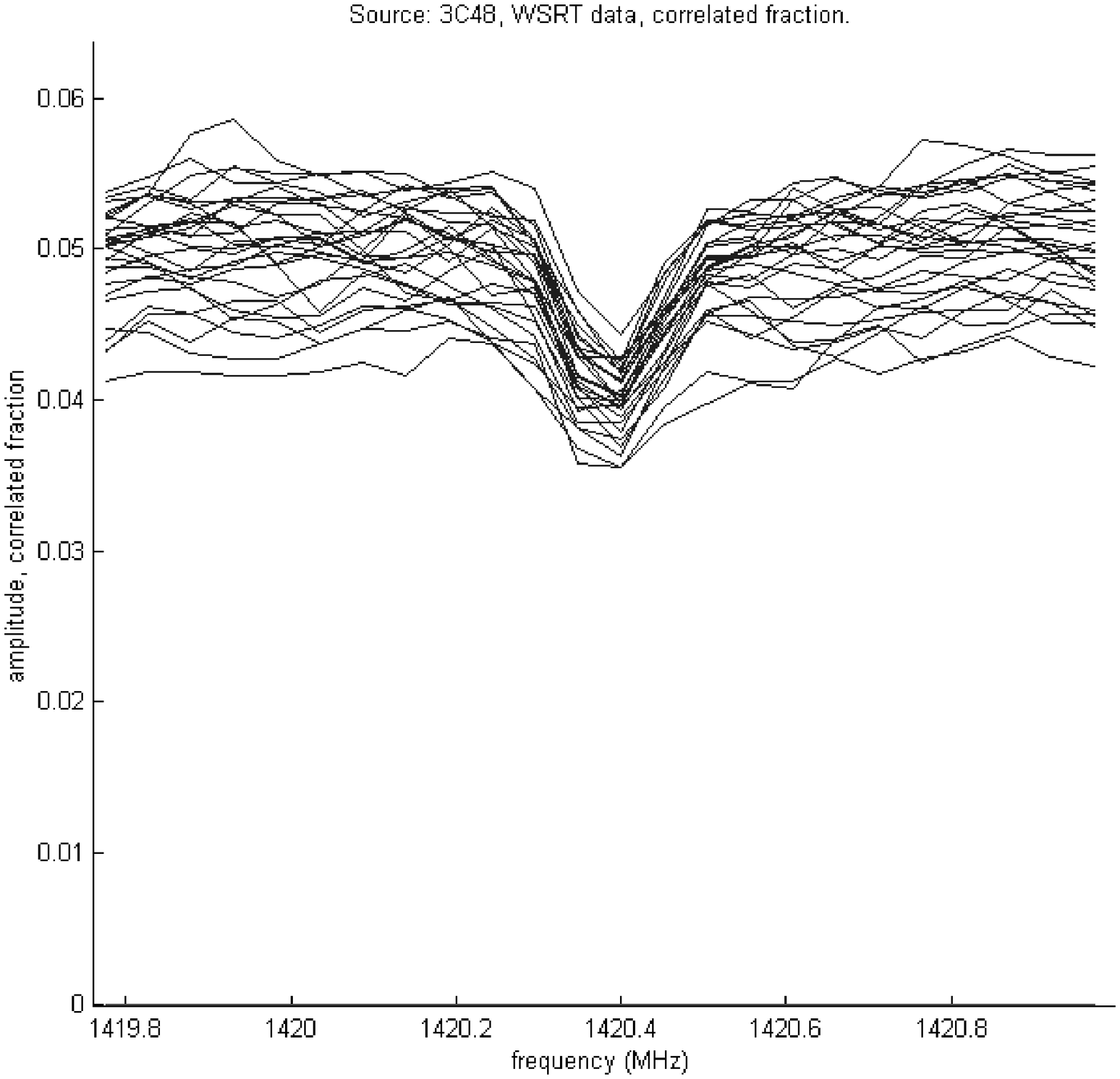,width=0.45\textwidth}}
    \hfill
    $(b)$
    \mbox{\psfig{figure=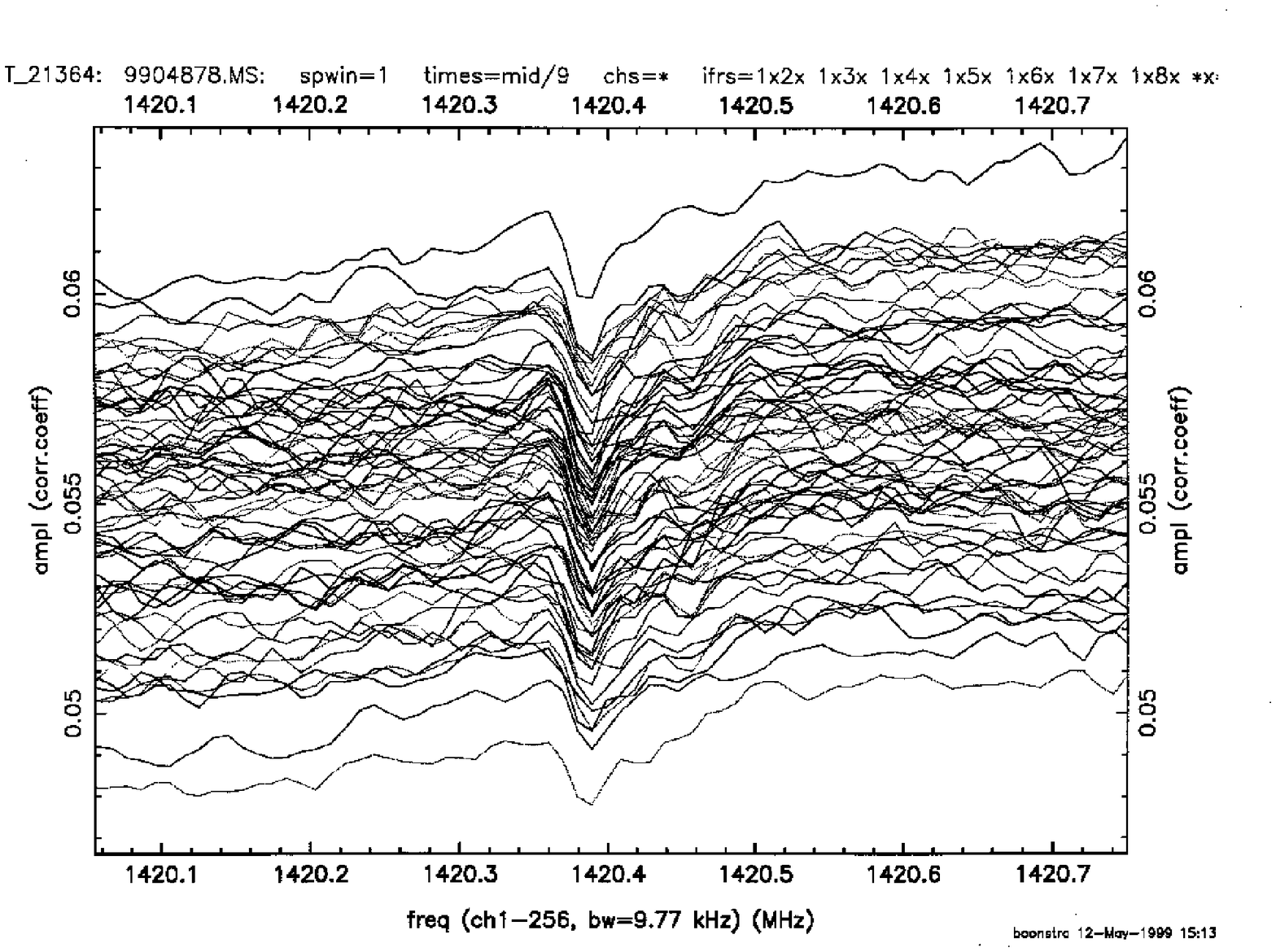,width=0.45\textwidth}}
  \caption{3C48 coherency function, magnitude (for all baselines). 
      $(a)$ NOEMI recording and off-line processing, 
      $(b)$ online WSRT processing by the DZB}
  \label{fig:coherency_sig14}
  \label{fig:dzb_zoom}
\end{figure}

    Figure \ref{fig:coherency_sig14}$(a)$ shows the magnitude of
    the coherency function 
    for all nonzero baselines as based on a NOEMI recording of a few
    seconds.  The coherency is around 5\%, and the spectral absorption
    at 1420.4~MHz shows up as a dip.
    We verified that the absorption line is statistically significant.
    For comparison we include the same spectral line as processed by the 
    WSRT DZB correlator in figure \ref{fig:dzb_zoom}$(b)$. 
    The values of the coherency are in good agreement  (differences are due
    to differences in processing bandwidths, observation times and 
    instrumental settings).

\begin{figure}
\begin{center}
    \mbox{\psfig{figure=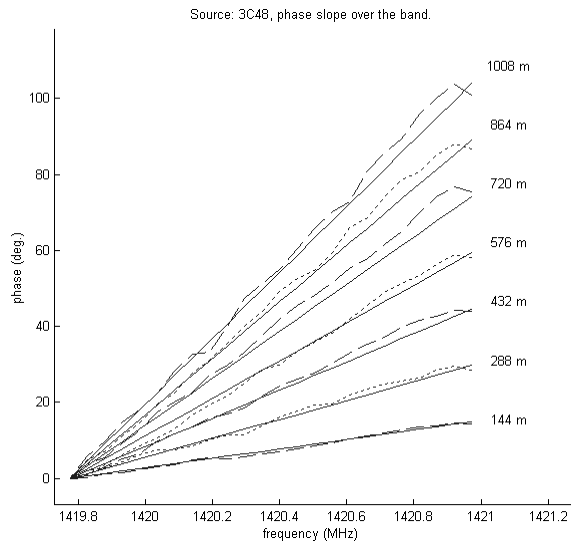,width=0.45\textwidth}}
\end{center}
    \caption{3C48 averaged coherency phase function vs.\ frequency, various
    baselines.}
    \label{figcoherency_phase}
\end{figure}

    To verify the phase behavior of the coherency we have computed the 
    unwrapped phase as a function of frequency.
    Note that the geometrical delay compensation and fringe corrections
    were not included in the recording.  Due to the
    narrowband processing, the delay offset $\tau_{ij}$
    of one channel with respect to another shows up as a frequency-dependent
    phase shift $e^{-j 2 \pi f \tau_{ij}}$ (the fringe),
    which will be the phase of $\rho_{ij}(f)$.  Here, $\tau_{ij}$ depends on
    the location of 3C48 and the baseline vector $\br_i-\br_j$ 
    between antenna $i$ and $j$, and is known.
    Figure \ref{figcoherency_phase} compares the observed phase differences
    (averaged over all identical baselines) to the computed phase, as a
    function of frequency and baseline length.  It is seen that
    the correspondence is very good. Note that for the shorter
    baselines we have more realizations so that their correspondence is
    better.

\subsection{3C48 absorption line with GSM interference}

    At this point we are ready to 
    demonstrate the performance of the sub-band detection and blanking method
    as presented in section \ref{sec:detection}.
    To this end, we have
    superimposed on the 3C48 data (at $1420$ MHz) another measurement file 
    containing
    GSM interference (at $905$ MHz), with the same bandwidth and for various 
    amplitude scalings of each file. 
    Although a bit artificial, the good linearity  of the WSRT system 
    implies that had a GSM signal been transmitted with a carrier 
    frequency of $1420$~MHz, then the measured data would be 
    the superposition of the two signals plus system noise. 
    The overlay allows us to verify the blanking performance for
    various mixtures of signal-to-interference power, since the clean
    data is now available as a reference and also the theoretical
    coherency is well known. 
    
\begin{figure}
\begin{center}
    \mbox{\psfig{figure=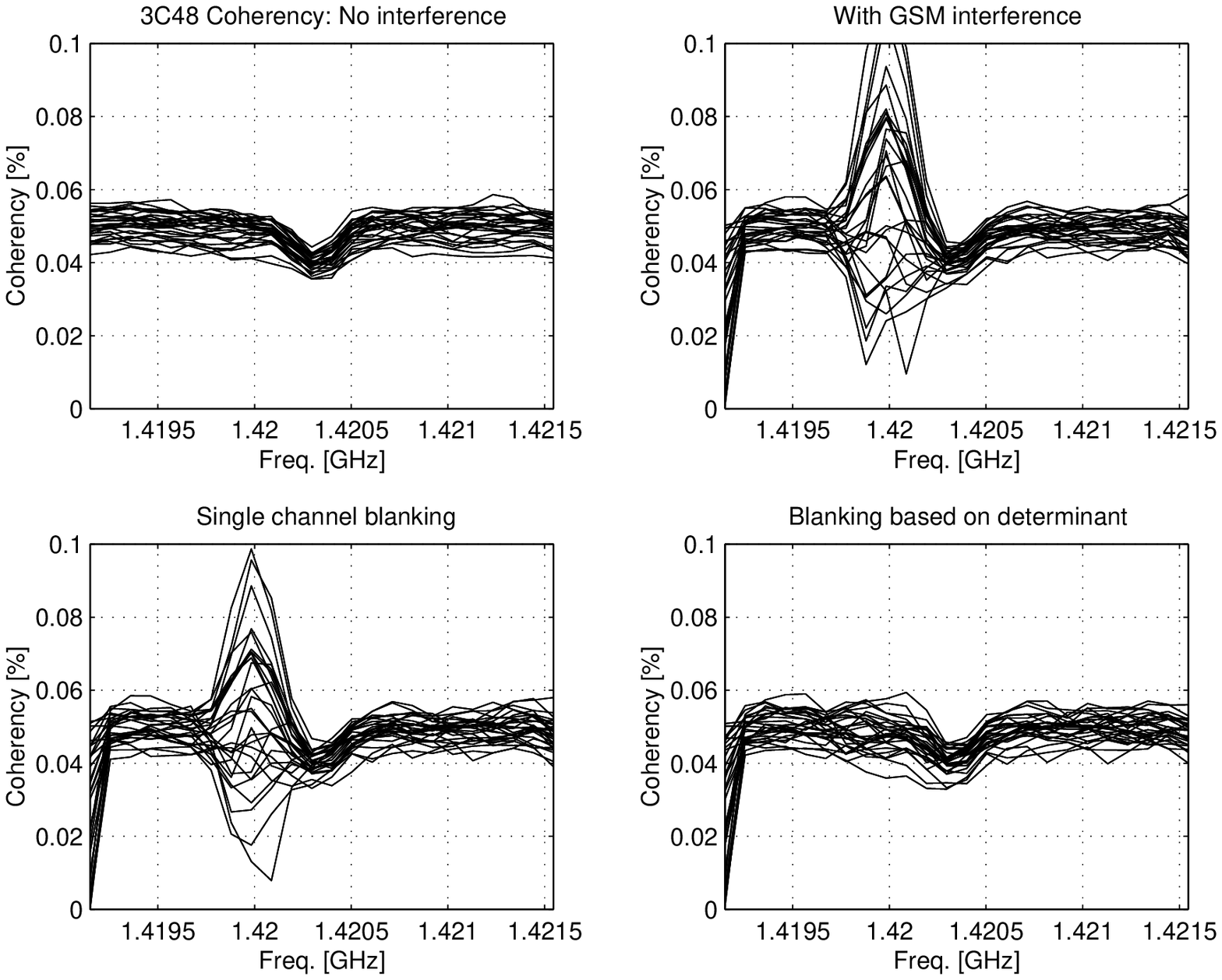,width=0.8\textwidth}}
\end{center}
 \caption{Magnitude of the coherency functions of 3C48 mixed with GSM
    interference. GSM data scaled by $0.1$.
    $(a)$ clean 3C48 data, 
    $(b)$ 3C48 mixed with GSM, 
    $(c)$ after single channel detection/blanking, 
    $(d)$ after multichannel subband detection/blanking}
  \label{fig:blank}
\end{figure}

    As described in section \ref{sec:detection}, the
    detection of an interferer in a specific time-frequency cell
    is based on the eigenvalues of the corresponding correlation matrix of
    the resulting mixture.  In this scheme, if one or more eigenvalues
    are above a threshold, then an interferer is detected and that data
    block is omitted.  However, to avoid the selection of
    the threshold based on a desired false alarm
    rate, we have chosen to simply throw away the worst 30 percent of the 
    data according to the value of the detector.
    We have computed the coherency of the clean, the contaminated and
    the blanked signals.  Figure \ref{fig:blank} shows the coherency
    functions over all baselines
    for a particular mixture of signals and interference:
    scaling the GSM data file by $0.1$ and the clean 3C48 data file by $0.9$.
    It is seen that $(a)$ the clean 3C48 spectrum shows the absorption line,
    which is $(b)$ completely masked when GSM interference is added.  After
    blanking, $(d)$ the absorption line is almost perfectly recovered.  For
    comparison we also included $(c)$ the results of blanking based on
    single channel power detection from channel 2 only, 
    without the sub-band decomposition. 
    The failure of this common way of single channel detection
    is clearly seen.  The reason is that the GSM signal was rather weak, 
    so that for single-channel wideband processing
    the probability of detection was quite low, even for a false alarm
    rate of up to $30\%$.

\begin{figure}
\begin{center}
    \mbox{\psfig{figure=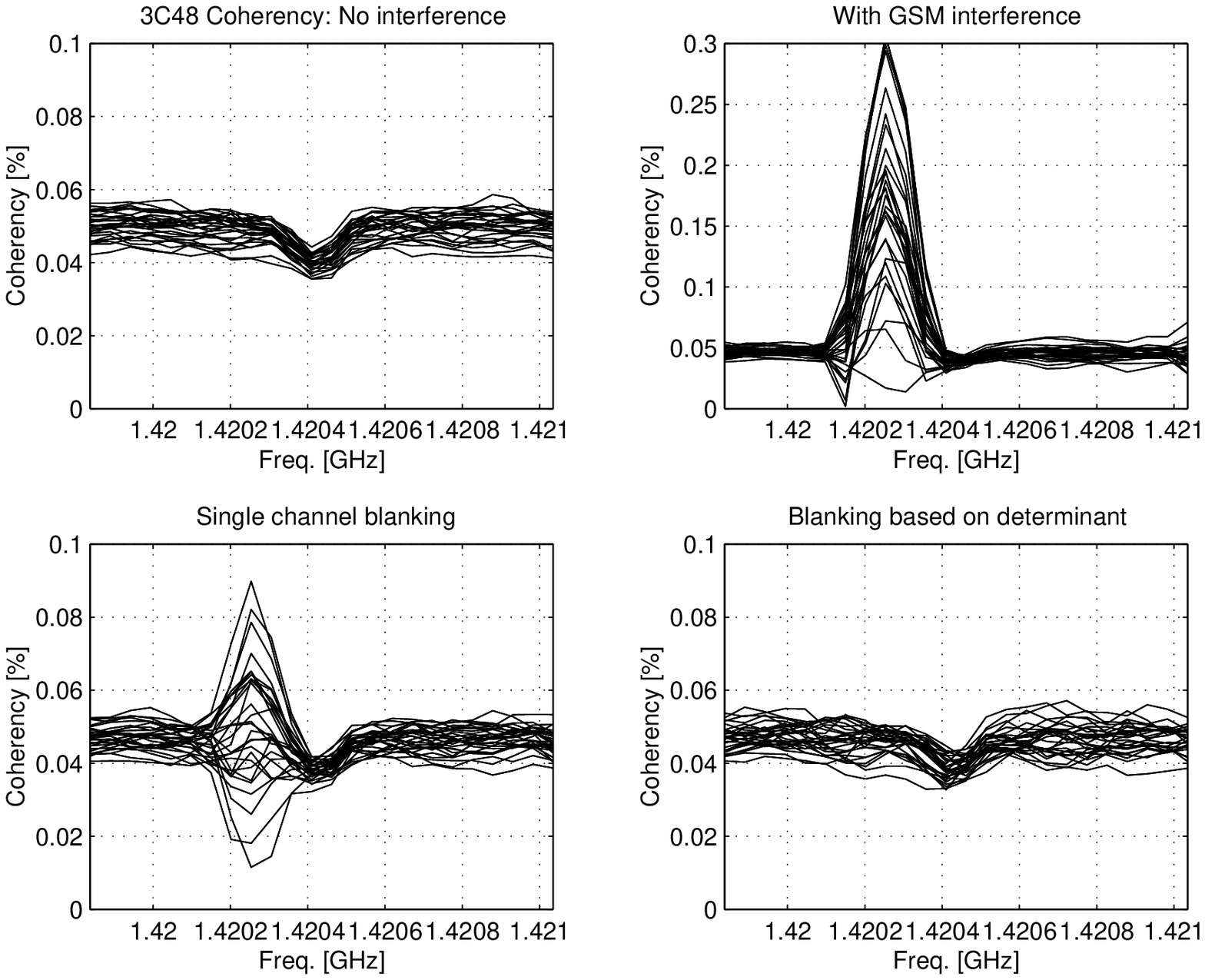,width=0.8\textwidth}}
\end{center}
 \caption{Magnitude of the coherency functions of 3C48 mixed with GSM
    interference. GSM data scaled by $0.5$.
    $(a)$ clean 3C48 data, 
    $(b)$ 3C48 mixed with GSM, 
    $(c)$ after single channel detection/blanking, 
    $(d)$ after multichannel subband detection/blanking}
  \label{fig:blank_strong}
\end{figure}

    To show the effect of interference power we have repeated the experiment 
    with the GSM data set weighted by a factor $0.5$.
    The stronger GSM interferer is now more easily detected
    and the resulting spectrum after blanking is better as seen in figure
    \ref{fig:blank_strong}.

\section{Conclusions} 					\label{sec:concl}

    In this paper,
    we considered various aspects of multichannel interference
    suppression for radio-astronomy. 
    It was shown that by sub-band processing we have access to the many
    narrow-band techniques available in array signal processing.
    We have demonstrated the benefits of multichannel 
    spatio-spectral blanking, both theoretically and on measured data. 
    The results are very pleasing.
    We have also discussed spatial filtering techniques and
    demonstrated how  they can be incorporated into the
    radio-astronomical measurement equation.

\acknowledgements

    Amir Leshem and A.J. Boonstra were supported by the NOEMI project 
    of the STW under contract no.~DEL77-4476.
    We would like to thank E.F.~Deprettere at TU Delft
    and our project partners at NFRA, 
    especially A.~van Ardenne, P.~Friedman, A.~Kokkeler, 
    J.~Noordam, and G.~Schoonderbeek, for the very useful collaboration.

\newpage

\newpage



\begin{figure}
\begin{center}
 \mbox{\psfig{file=leshem6.epsi,width=.8\textwidth}}
\end{center}
 \caption{Television broadcast}
 \label{fig:tvlingen}

\begin{center}
 \mbox{\psfig{file=leshem7.epsi,width=.8\textwidth}}
\end{center}
 \caption{GSM uplink}
 \label{fig:gsmup}
\end{figure}


\begin{figure}
\begin{center}
 \mbox{\psfig{file=leshem8.epsi,width=.8\textwidth}}
\end{center}
 \caption{GSM downlink}
 \label{fig:gsmdown}

\begin{center}
 \mbox{\psfig{file=leshem9.epsi,width=.8\textwidth}}
\end{center}
 \caption{Iridium downlink}
 \label{fig:iridium}
\end{figure}


\end{document}